\documentclass[acmsmall, nonacm]{acmart}

\usepackage{booktabs}
\usepackage{multirow}
\usepackage{threeparttable}
\usepackage{subcaption}
\newcommand{\para}[1]{%
  \smallskip
  \noindent\textbf{#1}%
}
\newif\ifcomments

\ifcomments
    \newcommand\han[1]{\textcolor{violet}{[Han: #1]}}
    \newcommand\cc[1]{\textcolor{blue}{[Chacha: #1]}}
    \newcommand{\chenhao}[1]{\textcolor{magenta}{\textsc{#1 ---CT}}}
    \else
    \newcommand\han[1]{}
    \newcommand\cc[1]{}
    \newcommand{\chenhao}[1]{}
\fi

\usepackage{xspace}
\usepackage{xparse}
\newcommand{\halone}{Human-alone\xspace}
\NewDocumentCommand{\hai}{o}{Human+AI\xspace\IfValueT{#1}{(Study #1)\xspace}}
\newcommand{\hensem}{Human-ensemble\xspace}
\newcommand{\haiensem}{Human+AI-ensemble\xspace}

\newcommand{\vp}{View Panel\xspace}
\newcommand{\cp}{Control Panel\xspace}
\newcommand{\ap}{Annotation Panel\xspace}

\usepackage{cleveref}
\crefname{figure}{Fig.}{Figs.}
\crefname{table}{Table}{Tables}

\newcommand{\tabcell}[1]{
  \begin{tabular}{@{}c@{}}
    #1
  \end{tabular}
} %
\newcommand{\cella}[3]{
  \begin{tabular}{@{}c@{}}
    #1\\$[#2, #3]$
  \end{tabular}
} %
\newcommand{\cellb}[5]{
  \begin{tabular}{@{}c@{}}
    #1\\$[#2, #3]$\\#4/#5
  \end{tabular}
} %

\usepackage{enumitem}

\AtBeginDocument{%
  \providecommand\BibTeX{{%
    \normalfont B\kern-0.5em{\scshape i\kern-0.25em b}\kern-0.8em\TeX}}}

\setcopyright{none}

\makeatletter
\def\@ACM@checkaffil{%
  \if@ACM@instpresent\else
  \ClassWarningNoLine{\@classname}{No institution present for an affiliation}%
  \fi
  \if@ACM@citypresent\else
  \ClassWarningNoLine{\@classname}{No city present for an affiliation}%
  \fi
}
\makeatother

\begin{document}

\title{Can Domain Experts Rely on AI Appropriately? A Case Study on AI-Assisted Prostate Cancer MRI Diagnosis}

\author{Chacha Chen}
\affiliation{%
  \institution{University of Chicago}
}
\email{chacha@uchicago.edu}

\author{Han Liu}
\affiliation{%
  \institution{University of Chicago}
}
\email{hanliu@uchicago.edu}

\author{Jiamin Yang}
\affiliation{%
  \institution{Toyota Technological Institute at Chicago}
}
\email{jiaminy@ttic.edu}

\author{Benjamin M. Mervak}
\affiliation{%
  \institution{University of Michigan}
}
\email{bmervak@med.umich.edu}

\author{Bora Kalaycioglu}
\affiliation{%
  \institution{University of Chicago}
}
\email{Bora.Kalaycioglu@bsd.uchicago.edu}

\author{Grace Lee}
\affiliation{%
  \institution{University of Chicago}
}
\email{glee@bsd.uchicago.edu}

\author{Emre Cakmakli}
\affiliation{%
  \institution{Bagcilar Training and Research Hospital}
}
\email{emre.cakmakli@std.yildiz.edu.tr}

\author{Matteo Bonatti}
\affiliation{%
  \institution{Hospital of Bolzano (SABES-ASDAA)}
}
\email{Matteo.bonatti@sabes.it}

\author{Sridhar Pudu}
\affiliation{%
  \institution{Radiology Associates of North Texas}
}
\email{spudu@radntx.com}

\author{Osman Kahraman}
\affiliation{%
  \institution{İstanbul Medipol University Hospital}
}
\email{osman.kahraman@medipol.edu.tr}

\author{Gül Gizem Pamuk}
\affiliation{%
  \institution{Bagcilar Training and Research Hospital}
}
\email{istanbul30@saglik.gov.tr}

\author{Aytekin Oto}
\affiliation{%
  \institution{University of Chicago}
}
\email{aoto@bsd.uchicago.edu}

\author{Aritrick Chatterjee}
\affiliation{%
  \institution{University of Chicago}
}
\email{aritrick@uchicago.edu}

\author{Chenhao Tan}
\affiliation{%
  \institution{University of Chicago}
}
\email{chenhao@uchicago.edu}

\renewcommand{\shortauthors}{Trovato and Tobin, et al.}

\begin{abstract}

Despite the growing interest in human-AI decision making, experimental studies with domain experts remain rare, largely due to the complexity of working with domain experts and the challenges in setting up realistic experiments.
In this work, we conduct an in-depth collaboration with radiologists in prostate cancer diagnosis based on MRI images.
Building on existing tools for teaching prostate cancer diagnosis, 
we develop an interface and conduct two experiments to study how AI assistance and performance feedback shape the decision making of domain experts.
In Study 1, clinicians were asked to provide an initial diagnosis (human), then view the AI’s prediction, and subsequently finalize their decision (human+AI).
In Study 2 (after a memory wash-out period), the same participants first received aggregated performance statistics from Study 1, specifically their own performance, the AI’s performance, and their human+AI performance, and then directly viewed the AI’s prediction before making their diagnosis (i.e., no independent initial diagnosis). 
These two workflows represent realistic ways that clinical AI tools might be used in practice, where the second study simulates a scenario where doctors can adjust their reliance and trust on AI based on prior performance feedback.
Our findings show that, while human+AI teams consistently outperform humans alone, they still underperform the AI due to under-reliance, similar to prior studies with crowdworkers.\han{Aritrick: ``I think this statement is best left for the discussion''. (referring to 'similar to...')}
Providing clinicians with performance feedback did not significantly improve the performance of human-AI teams, although showing AI decisions in advance nudges people to follow AI more. 
Meanwhile, we observe that the ensemble of human-AI teams can outperform AI alone, suggesting promising directions for human-AI collaboration.
Overall, our work highlights the prevalence and persistence of under-reliance, while demonstrating hope for complementary performance.
\end{abstract}



\maketitle

\section{Introduction}

AI holds promise for improving human decision making in a wide range of domains~\citep{lai2023towards,kleinberg2018human,reverberi2022experimental,scherer2019artificial,alon2023human}. 
Radiology is a representative example as 
AI outperforms or shows comparable performance with experts~\cite{hosny2018artificial,wu2019deep,rodriguez2019stand,rauschecker2020artificial,SahaBosmaTwilt2024,kromrey2024navigating,pauwels2021artificial,mckinney2020international}.
Rather than complete automation, there is growing consensus that AI's optimal role in the near future will serve as an assistance tool for human radiologists in clinical decision making~\cite{langlotz2019will, agarwal2023combining, norden2022ai, harvey2020fda}.
On the one hand, legal and regulatory challenges stand in the way of full automation. 
On the other hand, human AI collaboration has the potential to achieve \textit{complementary performance}, where human experts can leverage their contextual knowledge and expertise to correct AI mistakes in ways that could surpass either human or AI performance alone. 

However, the actual utility of integrating AI assistance tools in clinical settings remain poorly understood. 
In particular, very few studies examine the effectiveness of AI assistance in real clinical decision-making with domain experts~\cite{ouanes2024effectiveness, armando2023clinical}.
In this work, we conduct an in-depth collaboration with radiologists and focus on the case of prostate cancer diagnosis.
Prostate cancer diagnosis with\han{Aritrick: multiparametric} magnetic resonance imaging (MRI) remains one of the most difficult tasks for radiologists—even experienced ones—and inter-reader variability is high~\cite{de2014accuracy,chatterjee2025prospective}. Such complexity makes prostate MRI an ideal testbed for studying how AI assistance may complement human expertise. If AI can help reduce radiologists' mistakes here, it is plausible that similar technology could be effective in other radiology tasks as well.

We run human studies with domain experts to directly understand AI tool integration in radiology workflow, particularly for challenging diagnoses like prostate cancer. We investigate two key questions:
\begin{enumerate}[label=\textbf{Q\arabic*:}, itemsep=0.2em]
    \item \emph{Can AI-assistance help humans achieve higher diagnostic accuracy 
    than either human experts or AI systems alone?}
    \item \emph{How does AI-assistance shape human decision making beyond decision accuracy?}
\end{enumerate}

\begin{figure}[t]
    \centering
    \includegraphics[width=\linewidth]{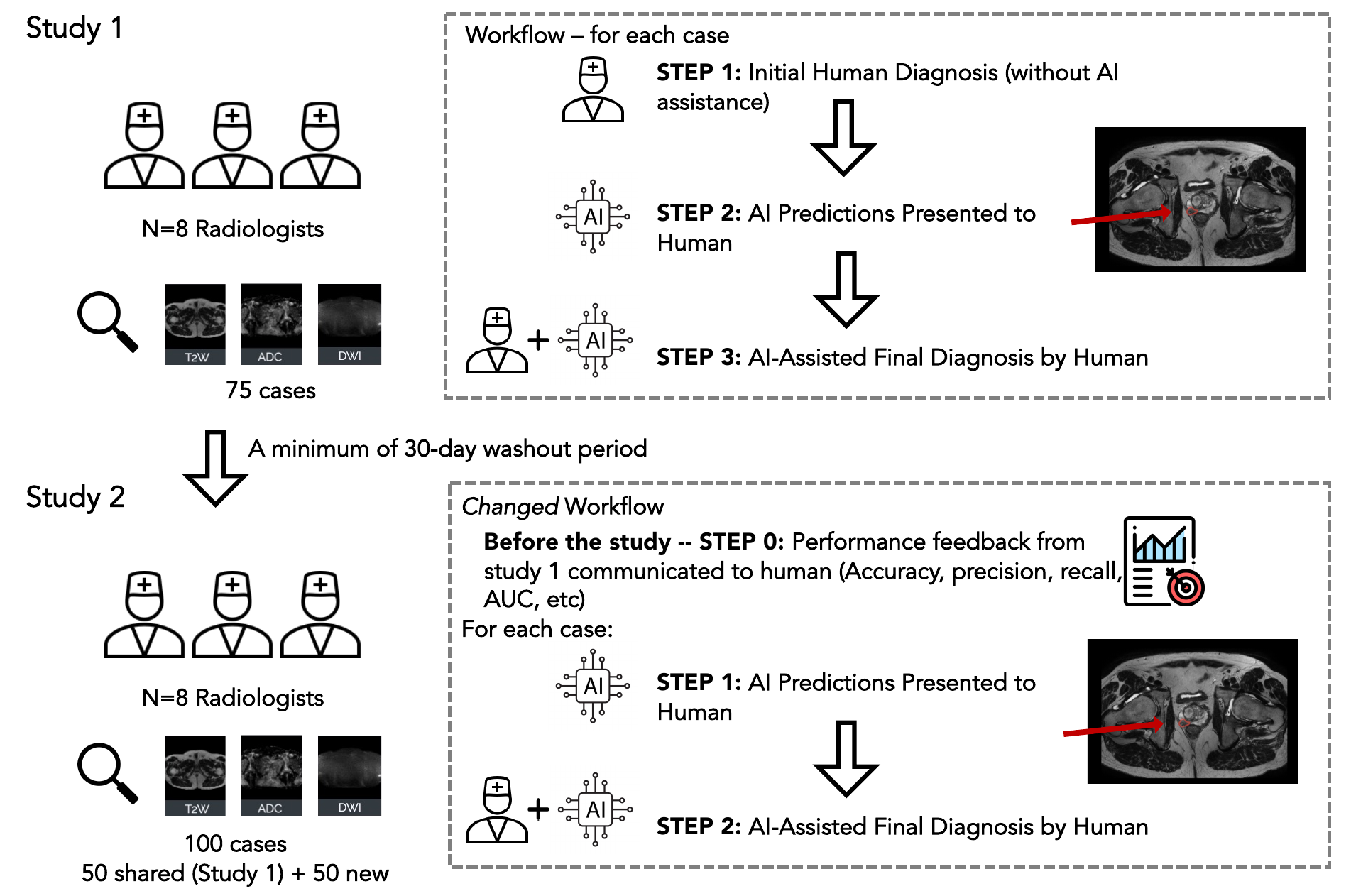}
    \caption{Overview of our experiments with radiologists.
    In study 1, participant radiologists (N=8) reviewed 75 cases in three steps: initial independent diagnosis, review of AI predictions, and final diagnosis. In study 2, we introduce performance feedback to communicate individual radiologist's performance collected from study 1 before the study. Then they reviewed 100 cases with direct AI assistance without independent diagnosis. 
    }
    \label{fig:study_overview}
\end{figure}

To answer these questions, we conducted pre-registered human subject experiments with domain experts, specifically board-certified radiologists (N=8), focusing on prostate cancer diagnosis with AI assistance.
We first trained a state-of-the-art AI model~\cite{isensee2021nnu} for prostate cancer detection from MRI scans.
The AI model is able to provide both diagnostic predictions and lesion annotation maps for positive cases as assistance for radiologists. 
To simulate real-world clinical practice, we designed and implemented two distinct workflows, see \cref{fig:study_overview} for an overview of the design of our human studies. 
Building on existing tools for teaching prostate cancer diagnosis, we also developed a web-based diagnostic platform that enables radiologists to review MRI scans and annotate suspicious cancer lesions seamlessly. 

In Study 1, radiologists each evaluated 75 cases in a three-step process.
For each case, they first made independent diagnoses, which helped us to establish baseline human performance.\han{Aritrick: For each case, they first made independent diagnoses using the PIRADS v2.1 guidelines to mimic standard of clinical care, which helps us to establish baseline human performance in the diagnosis of clinically significant cancers (Gleason $\geq$3+4).}
Then, they were shown the AI's predictions. In the final step, they are asked to finalize their decisions after reviewing AI predictions.
In Study 2, we introduced a novel element: before starting their evaluations, radiologists first received detailed individual performance feedback from Study 1, as shown in the screenshot in \cref{fig:feedback}.
This feedback included various metrics of their own performance, AI's performance, and their AI-assisted performance. 
To ensure engagement with this feedback, participants completed attention checks about their performance metrics before proceeding with new cases. This design allowed us to systematically examine how performance awareness influences radiologists' interaction with AI assistance.
Moreover, for each case diagnosis, AI assistance was provided directly to radiologists without them making independent diagnosis. 

These two distinct workflows represent common scenarios in the deployment of AI assistance tools in clinical practice and their evolution over time. 
Study 1 simulates an approach often regarded as responsible, as it allows radiologists to form independent opinions before consulting AI predictions. 
This approach may be particularly relevant during early deployments,
since radiologists may prefer minimal intervention to exercise caution.
Over time, the performance information will become available in a local scenario that retains the same distribution of doctors and patients as in the earlier integration of AI tools. 
Through the design of Study 2, we can investigate how both the timing of AI assistance and awareness of comparative performance metrics influence diagnostic accuracy and radiologists' integration of AI recommendations. \han{...diagnostic accuracy and the extent to which radiologists integrate AI recommendations / the way radiologists integrate AI recommendations?}

Our findings are consistent with prior studies on human-AI decision making.
Human+AI outperforms human alone, showcasing the positive utility of AI assistance.
However, Human+AI underperforms AI alone, largely driven by under-reliance.
Although performance feedback and upfront AI assistance nudged radiologists to incorporate AI predictions more frequently, we did not observe statistically significant improvements in metrics such as area under the receiver operating characteristic curve (AUROC/AUC) or accuracy.
We further investigate the effect of ensembling decisions.
A promising finding is that the majority vote of Human-AI teams can outperform AI alone, achieving complementary performance.
This observation points to exciting opportunities to identify insights into optimal ways to facilitate human-AI decision making.
\han{This observation highlights exciting opportunities to uncover insights into optimal strategies for facilitating human-AI decision making.}

To summarize, we make the following contributions:
\begin{itemize}[nosep]
    \item We conduct an in-depth collaboration with domain experts and design two experiments to study the effect of AI-assistance on expert decision making.
    \item We demonstrate that while human+AI outperforms human alone, they fall short of AI alone, similar to prior studies with crowdworkers.
    \item We present potential opportunities in leveraging the collective wisdom of human-AI teams.
\end{itemize}

\section{Related work}

\para{Human-AI decision making.}
There is a growing interest in the research community to augment human decision making with AI assistance~\cite{lai2023towards}. 
Typically, the tasks of interest are situated in high-stakes domains such as medicine, law, and finance, where AI-assisted decisions can have significant consequences. 
However, due to constraints related to resources and the simplicity of participant recruitment, the majority of empirical studies in this area are conducted with crowdworkers or laypeople without expertise.  
For instance, instead of involving real judges, researchers have explored recidivism prediction as a testbed for Human-AI decision making using crowdworkers~\cite{binns2018s, lai2019human, green2019disparate}.
Similarly, in the medical domain, experiments on disease diagnosis have been conducted with laypeople, such as students~\cite{lakkaraju2016interpretable}. 
In finance, studies have utilized crowdworkers for tasks like income prediction~\cite{zhang2020effect}, loan approval~\cite{green2019disparate}, and sales forecasting~\cite{dietvorst2015algorithm}. 
In some cases, researchers have substituted real-world tasks with entirely artificial ones to facilitate experimentation with crowdworkers, such as alien medicine recommendation~\cite{lage2019evaluation}.

While crowdworkers offer a convenient participant pool, it remains unclear if findings based on these populations generalize to domain experts in real cases. In our work, we work directly with domain experts.

\para{Human-AI decision making with experts in the clinical context.}
There have been several studies with healthcare professionals in the clinical context, but experiments focused on human-AI complementary performance remain limited.
While several studies have shown that AI assistance can improve diagnostic accuracy~\cite{steiner2018impact,sim2020deep,jain2021development,seah2021effect,mcduff2023towards}, the experts behavior in human-AI collaboration are underexamined.
Existing research also reveals complex performance trade-offs: some studies reveal important trade-offs, such as improved sensitivity at the cost of reduced specificity~\cite{kiani2020impact,park2019deep}.
Some studies explicitly demonstrated that the performance of human-AI performance falls short of AI alone~\cite{rajpurkar2020chexaid,kim2020changes}.
To the best of our knowledge, the only work that achieves complementary performance is \citet{steiner2018impact}, which demonstrated that algorithm-assisted pathologists outperformed both the algorithm and pathologists in detecting breast cancer metastasis. However, human specificity is 100\% on that task, suggesting a relatively easy task for domain experts.

In summary, human-AI decision making with domain experts, especially for complementary performance, remains underexplored. In light of this gap, our study aims to provide an in-depth analysis of both human+AI team performance and domain expert behavior in a difficult, real-world clinical setting.

\section{Methods}

\subsection{Dataset}

We used public data from the PI-CAI challenge\footnote{\url{https://pi-cai.grand-challenge.org/DATA/}} for training and testing.
The dataset originally contained 1500 cases, which we filtered down to 1411 cases by excluding cases from the same patients to avoid data leakage.
We ensure that all testing cases are biopsy-confirmed.
Our AI model was trained on 1211 cases, including 365 $(30.1\%)$ clinically significant prostate cancer (csPCa) cases. 
For study 1, the testing set includes 75 cases, of which 23 $(30.6\%)$ are csPCa.
Study 2 consists of 100 cases, with 32 $(32\%)$ being csPCa. 
For each patient case, we used T2-weighted (T2W), diffusion-weighted imaging (DWI), and apparent diffusion coefficient (ADC) sequences as inputs for both AI and human studies. 
50 cases were shared between study 1 and study 2, which allows us to directly compare performance metrics across both studies on this shared subset.

\para{Labels/annotations.} Case labels were obtained from three sources: biopsy-confirmed results (from systematic, magnetic resonance-guided biopsy, or prostatectomy), human-expert annotations, and AI-derived annotations~\cite{bosma2021annotation}. Out of the original 1500 cases, 1001 has biopsy confirmed case-level labels. 
Out of the 425 positive cases, 220 have human expert annotations, with the remaining annotated by AI. We prioritized human expert annotations when available, defaulting to AI annotations otherwise.
Ground truth case-level labels are approximately accurate, with 66.7\% (1001/1500) cases having biopsy results. Lesion-level annotations are less accurate due to the practical challenges of annotating all lesions in the large dataset. 
For all of our testing patient cases, case-level labels are derived from biopsy results. Lesion-level annotations are derived by experts (trained investigators and resident, supervised by expert radiologists), using all available clinical data. This includes MRI scans, diagnostic reports (radiology and pathology), and whole-mount prostatectomy specimens or other biopsy results when available.

\subsection{AI model \& performance}

We use the established nnU-Net model~\cite{isensee2021nnu,bosma2021annotation} as our AI model, trained from scratch with our own splits.  
We ensure that all testing examples have pathology groundtruth. 
Training examples have a mixture of different types of labels: pathology groundtruth, human expert labeled csPCa and delineation of the lesion area, and AI-labeled csPCa and lesion area~\cite{SahaBosmaTwilt2024}.
The AI standalone performance on the testing sets for both studies is shown in \cref{tab:main-results}.
The AI model achieves an AUROC of 0.910 in the training set, 0.730 and 0.790 respectively for the study 1 and study 2 testing set. 
Note that all testing examples have pathology groundtruth while as training sample have a mixture of pseudo labels.
For comprehensive details on the AI model's training configurations and performance metrics, please refer to \cref{appen:model}.%

\subsection{Human-AI Decision Making Interface}

\begin{figure}[t]
    \centering
    \begin{minipage}[b]{0.5\textwidth}
        \begin{subfigure}[b]{\textwidth}
            \includegraphics[width=\textwidth]{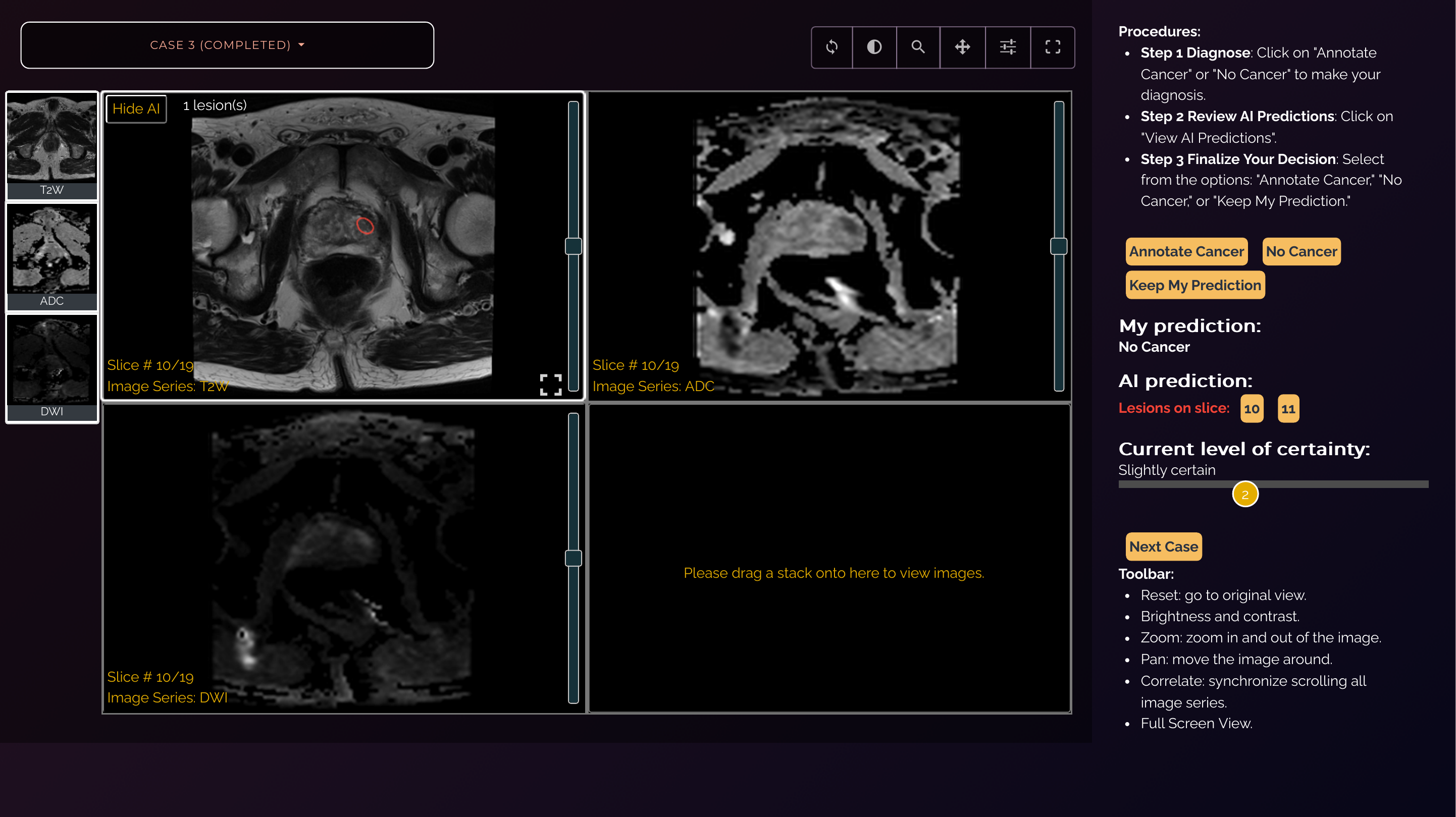}
            \caption{Patient case review interface.}
            \label{fig:study1_interface}
        \end{subfigure}
        \begin{subfigure}[b]{\textwidth}
            \includegraphics[width=\textwidth]{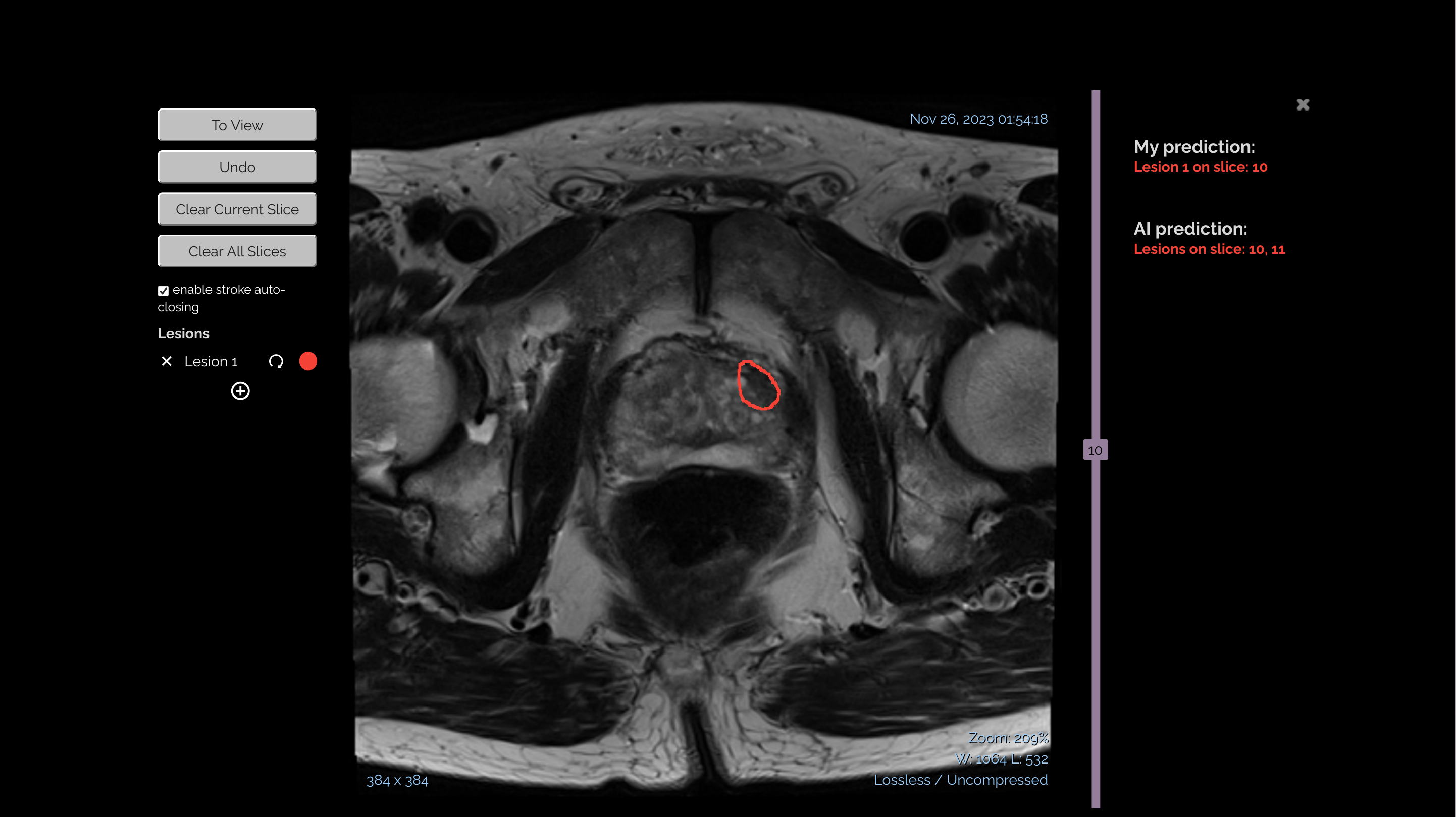}
            \caption{Lesion annotation panel.}
            \label{fig:annotate}
        \end{subfigure}
    \end{minipage}
    \begin{subfigure}[b]{0.465\textwidth}
        \includegraphics[width=\textwidth, trim={500 100 500 150}, clip]{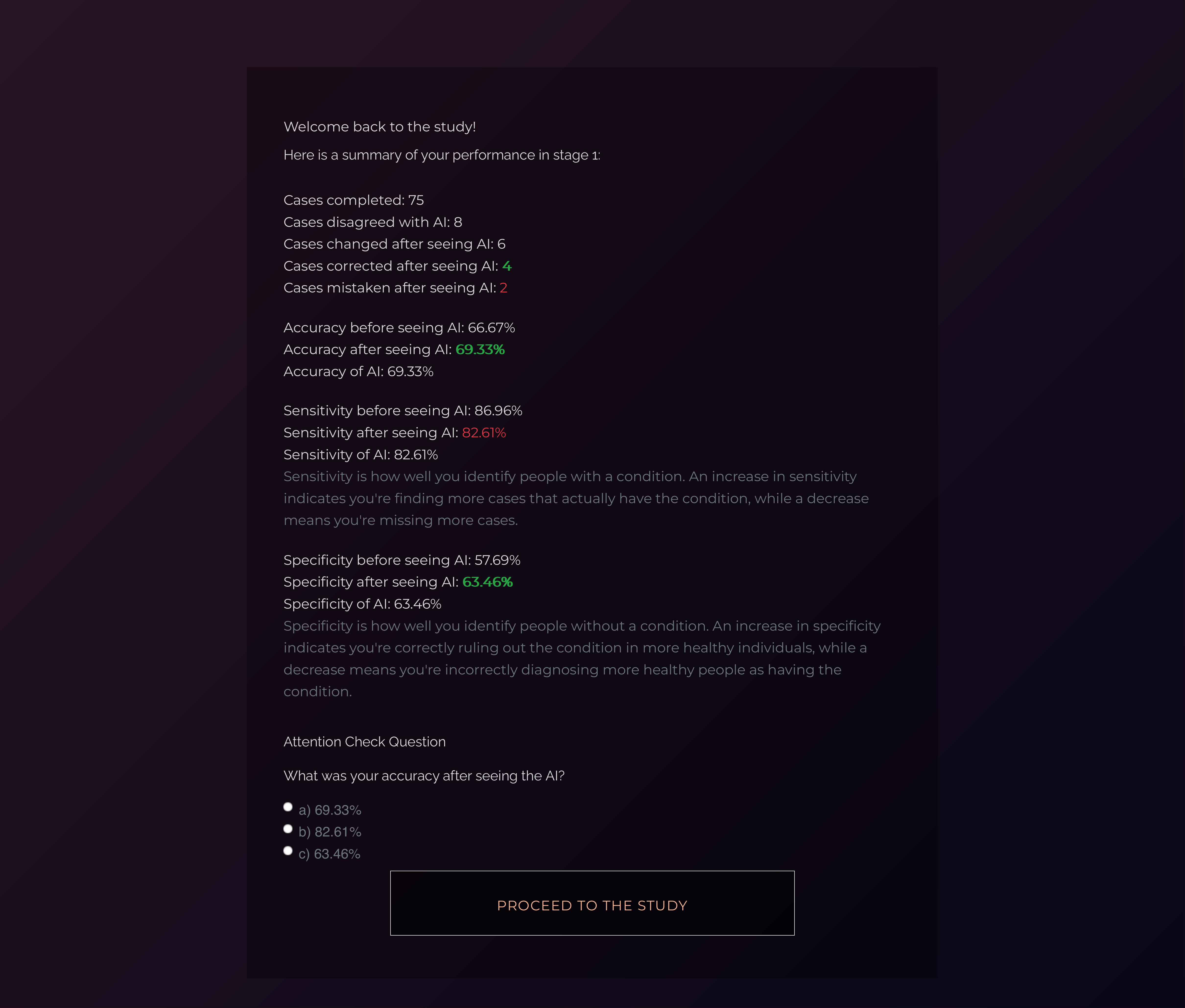}
        \caption{Performance feedback and attention check page.}
        \label{fig:feedback}
    \end{subfigure}
    \caption{
    Screenshots of the webapp interface for our human study.
    (a) \cref{fig:study1_interface} presents a user interface for patient case evaluation.
    An AI lesion prediction is highlighted with a red contour in the T2W sequence. On the right, the user's current prediction is shown as ``No Cancer," and they are at the stage of evaluating the AI prediction to make a final diagnosis. 
    (b) \cref{fig:annotate} shows the user interface of the \ap.
    The screenshot shows a current annotation of the user.
    The user can clear the annotation or add new annotations on the canvas.
    (c) \cref{fig:feedback} illustrates an example performance feedback page presented to a user before proceeding to Study 2. 
    The page provides a summary of the total number of cases, including counts of correct and incorrect cases, the number of decision changes influenced by AI advice, and whether those changes were correct or incorrect. It also highlights key performance metrics such as accuracy, sensitivity, and specificity, derived from Study 1.
    To ensure users review the information carefully, they are required to answer attention check questions.
    }
\end{figure}

We developed a webapp to conduct the human-study.
Participants can log in with their name and email.
They will see a consent page when they log in for the first time.
Once they give the consent, they will enter the study and see our study interface. A screenshot of the consent page can be found in appendix \cref{fig:consent-form}. Our human study is pre-registered and approved by the Institutional Review Board (IRB).

\para{Study interface.}
Our study interface has three major components: the \vp on the left, the \cp on the right, and the \ap as a pop-up in the center of the screen. 
The interface is shown in \cref{fig:study1_interface}.
In the \vp, we display three image sequences (T2W, ADC, BWI) from the MRI scans of the current case.
In the \cp, participants are informed about the current study (study 1 or 2) and provided with control buttons to make decisions or proceed to the next steps. 
Binary case-level AI predictions are also presented in this panel.  Participants make their own predictions by clicking the buttons (`Annotate Cancer" for positive cases and ``No Cancer'' for negative cases) and indicate their confidence level using a sliding bar.
If a participant believes the case is positive, they click the "Annotate Cancer" button, which triggers a pop-up window (\ap) displaying enlarged images from the T2W sequence of the current case, allowing participants to annotate the suspicious lesion areas. 
Participants can annotate any suspicious lesions by freely drawing on any image slice, using the sidebar to navigate between slices. 
The annotation interface is illustrated in \cref{fig:annotate}.

\para{Performance feedback.}
In Study 2, the first page after the login page will be the performance feedback page, as shown in \cref{fig:feedback}.
This page provides detailed individual feedback on their performance from Study 1.
The feedback includes both case counts and performance metrics. Specifically, we present the total number of cases completed by the participant, the number of cases where their prediction disagreed with the AI's prediction, and the number of times they changed their decision after viewing the AI's advice. Among these decision changes, we further highlight how many were correct and how many were mistaken after incorporating the AI's input.
For performance metrics, we provide accuracy, sensitivity, and specificity. These metrics are shown for the participant’s diagnoses before and after reviewing AI predictions, as well as for the AI’s performance alone. This breakdown allows participants to see the impact of the AI on their decision-making and compare their independent performance with AI.
At the bottom of the feedback page, we ask an attention check question to ensure participants review the information carefully.
The attention question is a single-answer multiple-choice question that asks for the value of one of the performance metrics displayed on the page.

\para{Exit survey.}
As the final step in both studies, participants are required to complete an exit survey. The survey for Study 1 collects demographic information and participants' opinions on AI. The survey for Study 2 gathers their thoughts on the performance feedback provided and revisits their opinions on AI. Screenshots of these surveys are included in the appendix \cref{fig:exit-survey-1} and \cref{fig:exit-survey-2}.

\subsection{Experimental Design}

To evaluate the effectiveness of AI assistance, we conduct two studies with practicing radiologists ($N=8$). An overview of our experimental workflow is shown in~\cref{fig:study_overview}.

Participant demographics, including experience levels, are detailed in Appendix~\ref{app:human_demographics}. Participants are recruited through interest forms distributed at the annual conference of RSNA (Radiological Society of North America), one of the largest radiology conferences in the world. 
We also use snowball recruiting, where participants refer colleagues and peers in their network. All participants are practicing radiologists and come from different regions (US and Europe), and all US-based participants are board-certified.

\para{Study conditions.}
Our experiments include three main conditions to evaluate radiologist performance:

\begin{itemize}[nosep]
    \item Human-only (Study 1): Independent diagnosis without AI assistance.
    \item \hai (Study 1): Diagnosis made after independent diagnosis and reviewing AI predictions.
    \item \hai (Study 2): Diagnosis made with AI predictions shown upfront, with prior feedback on individual performance metrics at the beginning of the study.
\end{itemize}

In Study 1, participants complete 75 test cases.
After logging in and signing the consent form, we provide a toy case to familiarize participants with the interface and workflow.
For each of the test cases, participants first make an independent diagnosis (human-only condition).
Then they review the AI prediction and annotations.
Participants have a chance to update and finalize their diagnosis before moving on to the next case (\hai condition for Study 1).

Between Study 1 and Study 2, we set a minimum memory wash-out period of 30 days to eliminate any recall effects. 
The actual period varies because participants complete the study at their own pace.

In Study 2, participants begin by reviewing a summary of their performance metrics from the \hai condition in Study 1. This feedback includes key metrics and interaction statistics to encourage reflection on their interaction with AI. To ensure engagement, participants answer an attention check question about the feedback before proceeding.
Study 2 consists of 100 cases, 50 randomly sampled from Study 1 and 50 new cases from a separate test pool.
Different from Study 1, AI predictions and annotations are shown upfront, and participants either accept the AI diagnosis or make modifications (\hai condition for Study 2).

Both studies conclude with an exit survey.

\subsection{Metrics and Statistical Testing Methods}

\para{Patient level metrics.}
We evaluate the performance using AUROC, accuracy, sensitivity/recall, specificity, negative predictive value (NPV), and positive predictive value (PPV)/precision, based on the predictions of Cancer vs. Non-Cancer for each case.
NPV is the proportion of cases predicted as Non-Cancer that are correctly classified. PPV/precision is the proportion of cases predicted as Cancer that are truly cancerous.

\para{Lesion level metrics.}
Note that lesion-level analysis focuses only on identified lesions (i.e., no true negatives), only accuracy, sensitivity, and PPV can be calculated at that level.
Prostate MRI consists of 3-D images, where lesions may span across multiple slices (images). For each 3-D connected lesion, we calculate lesion-level hits or misses based on a $10\%$ overlap between predicted annotations vs. groundtruth annotations, for both AI and human alike. 

\para{Statistical testing methods.}
We perform bootstrapped $z$-tests on the mean differences of metrics.
For each condition, bootstrapping is conducted by resampling with replacement over 10,000 iterations, using a sample size of 400 for population-level analysis and 50 for participant-level analysis.
We calculate the $95\%$ confidence intervals and $z$-statistics from the bootstrapped samples to conduct hypothesis testing.
Paired testing is performed when the data involve the same participants and cases; otherwise, unpaired testing is used.
We compute and report one-tailed $p$-values, applying a significance threshold of $\alpha=0.05$.

\begin{table}[t]
    \begin{center}
    \caption{Performance comparison between AI, Human, and Human+AI for identifying csPCa from MRI scans. 
    For each metric, the means, 95\% confidence intervals, and number of instances are reported. 
    The reported values and instance counts represent averages across eight radiologists.
    All confidence intervals are derived using bootstrap methods.  
    $p$-values are calculated using the bootstrap $z$-test with a significance threshold of $\alpha=0.05$.}
    \label{tab:main-results}
\resizebox{\textwidth}{!}{
    \begin{tabular}{lccccccc}
        \toprule
        & \multicolumn{7}{c}{Per-patient Analysis}  \\
        \midrule
        & \multicolumn{4}{c}{Study 1} &  \multicolumn{3}{c}{Study 2} \\
        \cmidrule(lr){2-5} \cmidrule(l){6-8}
        &  AI & Human  & Human+AI  & \tabcell{P (AI>Human)$^1$\\P (Human+AI>Human)\\P (AI>Human+AI)}  & AI & Human+AI   & \tabcell{P (Human+AI>Human)\\P (AI>Human+AI)} \\
        \midrule
        AUROC & \cella{0.730}{0.686}{0.772} & \cella{0.674}{0.627}{0.719} & \cella{0.701}{0.656}{0.746} & $0.023^*$/$0.033^*$/$0.131$ & \cella{0.790}{0.751}{0.829} & \cella{0.732}{0.689}{0.776} & $0.036^*$/$0.005^*$ \\ \midrule
        Accuracy & \cellb{69.3\%}{0.647}{0.738}{52}{75} & \cellb{63.2\%}{0.585}{0.677}{47}{75} & \cellb{66.2\%}{0.615}{0.708}{50}{75} & $0.013^*$/$0.009^*$/$0.103$ & \cellb{76.0\%}{0.718}{0.800}{76}{100} & \cellb{69.6\%}{0.650}{0.743}{70}{100} & $0.026^*$/$0.003^*$ \\ \midrule
        Sensitivity (Recall) & \cellb{82.6\%}{0.757}{0.891}{19}{23} & \cellb{78.3\%}{0.708}{0.853}{18}{23} & \cellb{80.4\%}{0.732}{0.874}{18}{23} & $0.171$/$0.207$/$0.299$ & \cellb{87.5\%}{0.815}{0.930}{28}{32} & \cellb{83.2\%}{0.765}{0.896}{27}{32} & $0.163$/$0.111$ \\ \midrule
        Specificity & \cellb{63.5\%}{0.577}{0.690}{33}{52} & \cellb{56.5\%}{0.507}{0.622}{29}{52} & \cellb{59.9\%}{0.542}{0.655}{31}{52} & $0.021^*$/$0.009^*$/$0.125$ & \cellb{70.6\%}{0.651}{0.759}{48}{68} & \cellb{63.2\%}{0.575}{0.691}{43}{68} & $0.052$/$0.006^*$ \\ \midrule
        NPV & \cellb{89.2\%}{0.847}{0.933}{33}{37} & \cellb{85.9\%}{0.803}{0.904}{29}{34} & \cellb{88.0\%}{0.826}{0.919}{31}{36} & $0.081$/$0.108$/$0.220$ & \cellb{92.3\%}{0.886}{0.958}{48}{52} & \cellb{89.3\%}{0.842}{0.932}{43}{48} & $0.159$/$0.052$ \\ \midrule
        PPV (Precision) & \cellb{50.0\%}{0.431}{0.569}{19}{38} & \cellb{44.7\%}{0.378}{0.509}{18}{41} & \cellb{47.1\%}{0.403}{0.537}{18}{39} & $0.014^*$/$0.012^*$/$0.105$ & \cellb{58.3\%}{0.514}{0.654}{28}{48} & \cellb{51.9\%}{0.447}{0.585}{27}{52} & $0.066$/$0.003^*$ \\
        \midrule
        & \multicolumn{7}{c}{Per-lesion Analysis$^2$}  \\\midrule
        & \multicolumn{4}{c}{Study 1$^3$} &  \multicolumn{3}{c}{Study 2} \\
        \cmidrule(lr){2-5} \cmidrule(l){6-8}
        & AI & Human  & Human+AI  & \tabcell{P (AI>Human)\\P (Human+AI>Human)\\P (AI>Human+AI)}  & AI & Human+AI   & \tabcell{P (Human+AI>Human)\\P (AI>Human+AI)}\\
        \midrule
        Accuracy & \cellb{35.4\%}{0.307}{0.403}{17}{48} & \cellb{25.7\%}{0.212}{0.297}{13}{53} & \cellb{28.5\%}{0.240}{0.330}{15}{51} & $0.001^*$/$0.168$/$0.019^*$ & \cellb{36.9\%}{0.323}{0.417}{24}{65} & \cellb{33.8\%}{0.292}{0.385}{22}{66} & $0.005^*$/$0.170$ \\ \midrule
        Sensitivity (Recall) & \cellb{73.9\%}{0.675}{0.800}{17}{23} & \cellb{58.4\%}{0.509}{0.658}{13}{23} & \cellb{63.4\%}{0.561}{0.706}{15}{23} & $0.001^*$/$0.176$/$0.015^*$ & \cellb{72.7\%}{0.665}{0.787}{24}{33} & \cellb{67.4\%}{0.608}{0.737}{22}{33} & $0.036^*$/$0.121$ \\ \midrule
        PPV (Precision) & \cellb{40.5\%}{0.353}{0.456}{17}{42} & \cellb{31.5\%}{0.261}{0.361}{13}{43} & \cellb{34.4\%}{0.290}{0.394}{15}{43} & $0.005^*$/$0.202$/$0.045^*$ & \cellb{42.9\%}{0.377}{0.482}{24}{56} & \cellb{40.6\%}{0.350}{0.456}{22}{55} & $0.006^*$/$0.247$ \\
        \bottomrule
    \end{tabular}
    }
    \end{center}
    \begin{flushleft}
    \small{$^1$$p$-values compare the performance of different conditions using bootstrap $z$-test. In Study 1, a paired test is conducted on 75 cases, where each case is evaluated by both Human Alone and Human+AI. In Study 2, an unpaired test is performed, comparing the performance on 75 Human Alone cases and 100 Human+AI cases.}\\
    \small{$^2$Note that the lesion-level analysis should be interpreted with caution compared to the per-patient analysis. Since lesion-level analysis excludes true negatives (TNs), we only calculate metrics that do not rely on TNs, i.e. accuracy, sensitivity and PPV.\\
    $^3$For study 1 lesion-level human results, one radiologist's results were excluded because they used our annotation tool incorrectly.}
    \end{flushleft}
\end{table}

\section{Results}

We organize our findings into two parts: 1) the effect of AI assistance on the performance of human-AI decision making; 2) how AI assistance changes behavioral patterns such as reliance and decision efficiency.
Overall, for (1), we observe a performance trend in order of \textbf{Human alone < Human+AI < AI}, with occasional instances of individual radiologists achieving complementary performance. It is also worth noting the ensemble of human+AI could outperform AI, i.e., complementary performance. 
For (2), we find that the different workflow does not significantly impact human performance. Radiologists are generally reluctant to adopt AI suggestions after making their own diagnosis. In contrast, providing upfront AI input increases the adoption of AI advice among experts. However, under-reliance on AI persists, preventing human+AI team from achieving complementary performance. 

\subsection{Performance of Human vs. AI vs. Human+AI Team (Q1)}
We evaluate both the baseline performance of humans and their performance after receiving AI assistance.
\cref{tab:main-results} presents an overview of performance metrics from both studies, including per-patient and per-lesion results. 

\begin{figure}[t]
    \centering
    \includegraphics[width=0.8\linewidth]{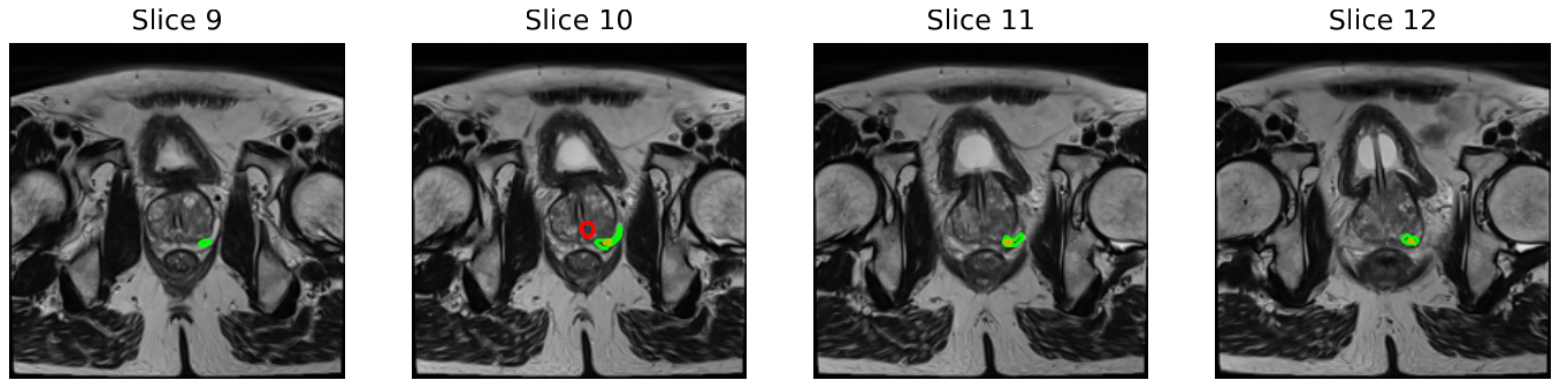}
    \caption{An example of lesion-level annotation comparing human experts (red contour), AI (yellow contour), and expert annotation from the dataset (green contour). In this case, the AI successfully detected a lesion which corresponded to a clinically significant prostate cancer in the dataset; our human radiologist did not identify this lesion, and instead annotated a lesion in the transition zone.}
    \label{fig:human-wrong-lesion}
\end{figure}

\para{\halone < AI.}
The workflow of Study 1 allows us to compare the baseline performance of humans and AI on the same set of patient cases. 
As shown in \cref{tab:main-results}, 
AI consistently outperforms humans across most metrics, with statistically significant advantages in AUROC, accuracy, specificity, and PPV/precision($p<0.05$).
At the lesion level, the AI also shows significant gains in accuracy, sensitivity, and PPV. Moreover, we find that for identified positive lesions, AI is less likely to miss the biopsy confirmed lesions, compared with human radiologists. \cref{fig:human-wrong-lesion} provides an example of this.
These findings suggest that the AI is better than human radiologists in predicting csPCa, especially in identifying true negative cases and true positive lesions.

\begin{table}[t]
    \begin{center}
    \caption{Performance comparison between AI, Human, and Human+AI (study 1 and 2) for the common 50-case subset. 
    $p$-values are calculated using the bootstrap $Z$-test, with a significance threshold of $\alpha=0.05$.
    }
    \label{tab:common-cases}
\resizebox{\textwidth}{!}{
    \begin{tabular}{lcccccc}
        \toprule
        && \multicolumn{3}{c}{Study 1} &  \multicolumn{2}{c}{Study 2} \\
        \cmidrule(lr){3-5} \cmidrule(l){6-7}
        & AI & Human & Human+AI & \tabcell{P(AI>Human)\\P(Human+AI>Human)\\P(AI>Human+AI)} & Human+AI & \tabcell{P(Human+AI>Human)\\P(AI>Human+AI)} \\
        \midrule
        AUROC & \cella{0.763}{0.727}{0.797} & \cella{0.675}{0.630}{0.719} & \cella{0.711}{0.668}{0.752} & $0.001^*$/$0.004^*$/$0.018^*$ & \cella{0.708}{0.666}{0.748} & $0.074$/$0.005^*$ \\ \midrule
        Accuracy & \cellb{70.0\%}{0.657}{0.745}{35}{50} & \cellb{62.5\%}{0.578}{0.672}{31}{50} & \cellb{65.7\%}{0.610}{0.703}{33}{50} & $0.003^*$/$0.002^*$/$0.045^*$ & \cellb{64.7\%}{0.600}{0.693}{32}{50} & $0.157$/$0.011^*$ \\ \midrule
        Sensitivity (Recall) & \cellb{93.8\%}{0.892}{0.976}{15}{16} & \cellb{81.2\%}{0.741}{0.878}{13}{16} & \cellb{85.9\%}{0.797}{0.917}{14}{16} & $0.001^*$/$0.028^*$/$0.017^*$ & \cellb{87.5\%}{0.815}{0.929}{14}{16} & $0.041^*$/$0.021^*$ \\ \midrule
        Specificity & \cellb{58.8\%}{0.530}{0.646}{20}{34} & \cellb{53.7\%}{0.477}{0.595}{18}{34} & \cellb{56.2\%}{0.504}{0.620}{19}{34} & $0.068$/$0.017^*$/$0.216$ & \cellb{54.0\%}{0.482}{0.599}{18}{34} & $0.450$/$0.058$ \\ \midrule
        NPV & \cellb{95.2\%}{0.918}{0.982}{20}{21} & \cellb{87.0\%}{0.804}{0.909}{18}{21} & \cellb{90.8\%}{0.846}{0.938}{19}{21} & $0.001^*$/$0.015^*$/$0.015^*$ & \cellb{91.4\%}{0.854}{0.945}{18}{20} & $0.043^*$/$0.014^*$ \\ \midrule
        PPV (Precision) & \cellb{51.7\%}{0.453}{0.581}{15}{29} & \cellb{45.5\%}{0.389}{0.517}{13}{29} & \cellb{48.2\%}{0.416}{0.545}{14}{29} & $0.003^*$/$0.003^*$/$0.045^*$ & \cellb{47.4\%}{0.410}{0.537}{14}{30} & $0.136$/$0.011^*$ \\
        \bottomrule
    \end{tabular}
    }
    \end{center}
\end{table}

\para{\halone < \hai.}
In Study 1, human+AI outperformed human radiologists alone, with statistical significance in AUROC, accuracy, specificity, and PPV/precision ($p<0.05$), as shown in \cref{tab:main-results}. This highlights the potential positive utility of AI assistance.

While study 2 did not include a direct human-alone baseline, we conducted two statistical analysis to evaluate the impact of AI assistance. First, we performed an unpaired statistical test, comparing human-alone performance from Study 1 (75 cases) against human+AI performance from Study 2 (100 cases). This analysis shows statistically significant improvements in both AUROC and accuracy, from \cref{tab:main-results}. Second, to further validate these findings with a common set of patient cases, we investigate specifically the 50 common cases shared between both studies to perform a paired statistical analysis. By referencing the human-alone performance from Study 1 on these exact same cases, we found that human+AI  outperformed human-alone in both studies, as shown in \cref{tab:common-cases}.

Overall, our findings suggest that AI assistance consistently improves radiologists' performance.

\para{\hai < AI.}
Although the Human + AI team outperforms humans alone, it consistently underperforms AI alone in AUROC, accuracy, specificity, and PPV/precision ($p<0.05$) in Study 2, while showing no significant evidence of inferiority to AI in Study 1.
This trend becomes more salient when focusing on the common 50-case subset, as shown in \cref{tab:common-cases}, where all metrics except specificity show statistically significant differences in both studies.
This is somewhat justified, as human radiologists in practice tend to be more cautious to avoid missing any suspicious cases (i.e., identifying true negative cases). They are inclined to send suspicious cases for biopsy.  
For lesion level analysis, it is more prominent that AI outperformed \hai in identifying positive lesions, with statistical significance in accuracy, sensitivity, and precision in Study 1.

\begin{figure}[t]
  \centering
  \begin{subfigure}[b]{0.4\textwidth}
    \centering
    \includegraphics[width=\textwidth]{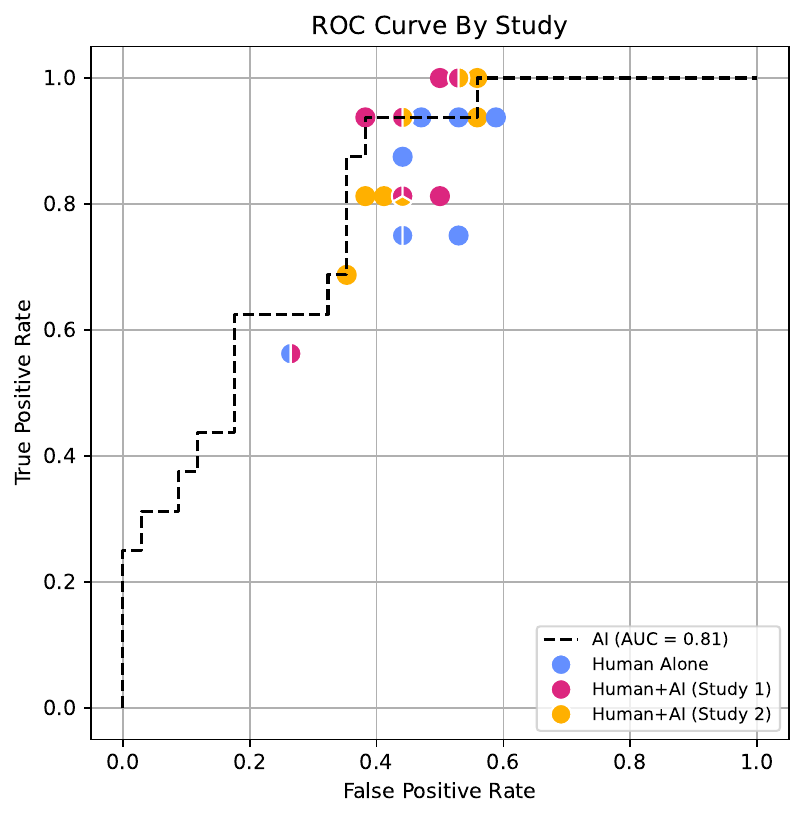}
    \caption{ROC Curve}
  \end{subfigure}
  \begin{subfigure}[b]{0.4\textwidth}
    \centering
    \includegraphics[width=\textwidth]{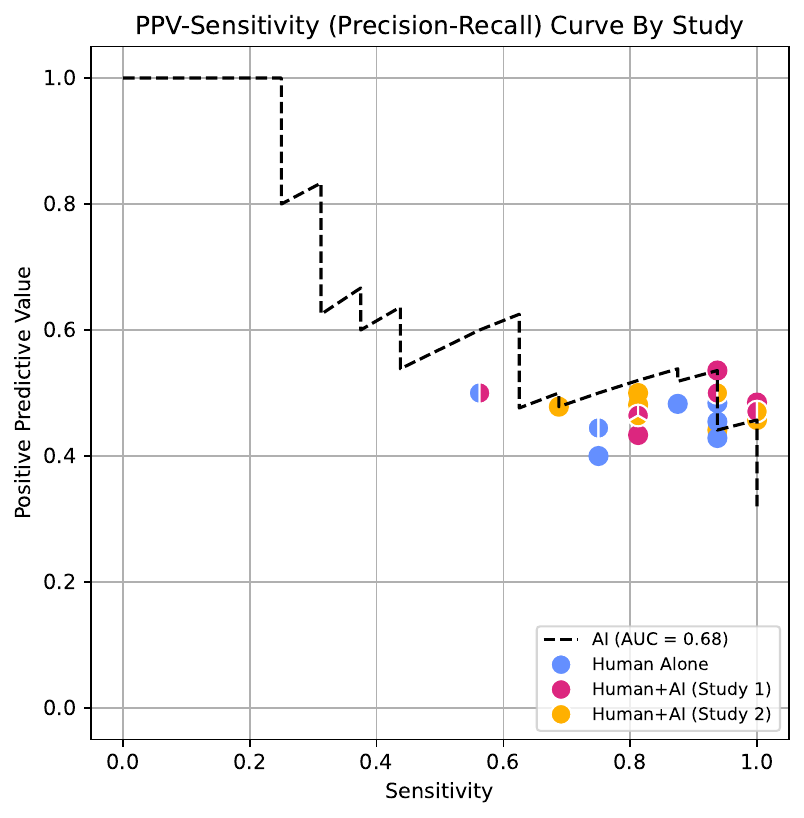}
    \caption{PR Curve}
  \end{subfigure}
  
  \caption{Individual radiologists performance compared with the AI model. The model achieves higher performance than all of the radiologists without AI assistance (blue dots). However, with AI assistance, some individual radiologists outperformed the AI model (red and orange dots that are above the curve).
  }
  \label{fig:two_side_by_side}
\end{figure}

\para{Individual human radiologists can occasionally achieve complementary performance.}
In the common cases between Study 1 and Study 2, we evaluate individual radiologists and AI-assisted radiologists against AI model using both receiver operating characteristic (ROC) and precision-recall (PR) curves. 
As shown in~\cref{fig:two_side_by_side}, and consistent with prior discussions, the AI curve generally outperforms individual radiologists (represented by blue dots).
Additionally, AI-assisted radiologists in both studies (red and orange dots) are generally positioned above individual radiologists (blue dots) in both figures, indicating that AI assistance helps improve radiologists’ performance.
We highlight that there are cases where AI-assisted radiologists outperform the AI curve, as shown by the red and orange dots above the AI curve. This is a promising finding as it suggests that AI assistance could augment human to achieve complementary performance (Human+AI > human and Human+AI > AI).

\begin{table}[t]
    \begin{center}
    \caption{Performance comparison between AI, Human, Human+AI, Human-ensemble, and Human+AI-ensemble for identifying csPCa from MRI scans.
    For each metric, the means, 95\% confidence intervals, and number of instances are reported. 
    The reported values and instance counts represent averages across eight radiologists.
    All confidence intervals are derived using bootstrap methods.  
    $p$-values are calculated using the bootstrap $z$-test with a significance threshold of $\alpha=0.05$.}
    \label{tab:ensemble}
\resizebox{\textwidth}{!}{
    \begin{tabular}{lcccccccccc}
        \toprule
        & \multicolumn{6}{c}{Study 1} & \multicolumn{4}{c}{Study 2}\\
        \cmidrule(lr){2-7} \cmidrule(l){8-11}
        & AI & Human & Human-ensemble & Human+AI & H+AI ensemble & \tabcell{P (\hensem>Human)\\P (H+AI ensemble>AI)} & AI & Human+AI & H+AI ensemble & P (H+AI ensemble>AI)  \\ \midrule
        AUROC & \cella{0.730}{0.686}{0.772} & \cella{0.674}{0.627}{0.719} & \cella{0.721}{0.677}{0.764} & \cella{0.701}{0.656}{0.746} & \cella{0.771}{0.730}{0.811} & $0.013^*$/$0.034^*$ & \cella{0.790}{0.751}{0.829} & \cella{0.732}{0.689}{0.776} & \cella{0.783}{0.743}{0.823} & $0.323$ \\ \midrule
        Accuracy & \cellb{69.3\%}{0.647}{0.738}{52}{75} & \cellb{63.2\%}{0.585}{0.677}{47}{75} & \cellb{68.0\%}{0.635}{0.725}{51}{75} & \cellb{66.2\%}{0.615}{0.708}{50}{75} & \cellb{73.3\%}{0.690}{0.777}{55}{75} & $0.009^*$/$0.046^*$ & \cellb{76.0\%}{0.718}{0.800}{76}{100} & \cellb{69.6\%}{0.650}{0.743}{70}{100} & \cellb{75.0\%}{0.708}{0.792}{75}{100} & $0.277$ \\ \midrule
        Sensitivity (Recall) & \cellb{82.6\%}{0.757}{0.891}{19}{23} & \cellb{78.3\%}{0.708}{0.853}{18}{23} & \cellb{82.6\%}{0.758}{0.891}{19}{23} & \cellb{80.4\%}{0.732}{0.874}{18}{23} & \cellb{87.0\%}{0.805}{0.926}{20}{23} & $0.098$/$0.090$ & \cellb{87.5\%}{0.815}{0.930}{28}{32} & \cellb{83.2\%}{0.765}{0.896}{27}{32} & \cellb{87.5\%}{0.816}{0.930}{28}{32} & $0.496$ \\ \midrule
        Specificity & \cellb{63.5\%}{0.577}{0.690}{33}{52} & \cellb{56.5\%}{0.507}{0.622}{29}{52} & \cellb{61.5\%}{0.559}{0.672}{32}{52} & \cellb{59.9\%}{0.542}{0.655}{31}{52} & \cellb{67.3\%}{0.619}{0.728}{35}{52} & $0.025^*$/$0.109$ & \cellb{70.6\%}{0.651}{0.759}{48}{68} & \cellb{63.2\%}{0.575}{0.691}{43}{68} & \cellb{69.1\%}{0.636}{0.747}{47}{68} & $0.255$ \\ \midrule
        NPV & \cellb{89.2\%}{0.847}{0.933}{33}{37} & \cellb{85.9\%}{0.803}{0.904}{29}{34} & \cellb{88.9\%}{0.843}{0.932}{32}{36} & \cellb{88.0\%}{0.826}{0.919}{31}{36} & \cellb{92.1\%}{0.882}{0.956}{35}{38} & $0.043^*$/$0.059$ & \cellb{92.3\%}{0.886}{0.958}{48}{52} & \cellb{89.3\%}{0.842}{0.932}{43}{48} & \cellb{92.2\%}{0.883}{0.957}{47}{51} & $0.457$ \\ \midrule
        PPV (Precision) & \cellb{50.0\%}{0.431}{0.569}{19}{38} & \cellb{44.7\%}{0.378}{0.509}{18}{41} & \cellb{48.7\%}{0.420}{0.555}{19}{39} & \cellb{47.1\%}{0.403}{0.537}{18}{39} & \cellb{54.1\%}{0.471}{0.610}{20}{37} & $0.010^*$/$0.049^*$ & \cellb{58.3\%}{0.514}{0.654}{28}{48} & \cellb{51.9\%}{0.447}{0.585}{27}{52} & \cellb{57.1\%}{0.502}{0.642}{28}{49} & $0.268$ \\
        \bottomrule
    \end{tabular}
    }
    \end{center}
\end{table}

\begin{figure}[ht]
    \centering
    \includegraphics[width=0.5\textwidth]{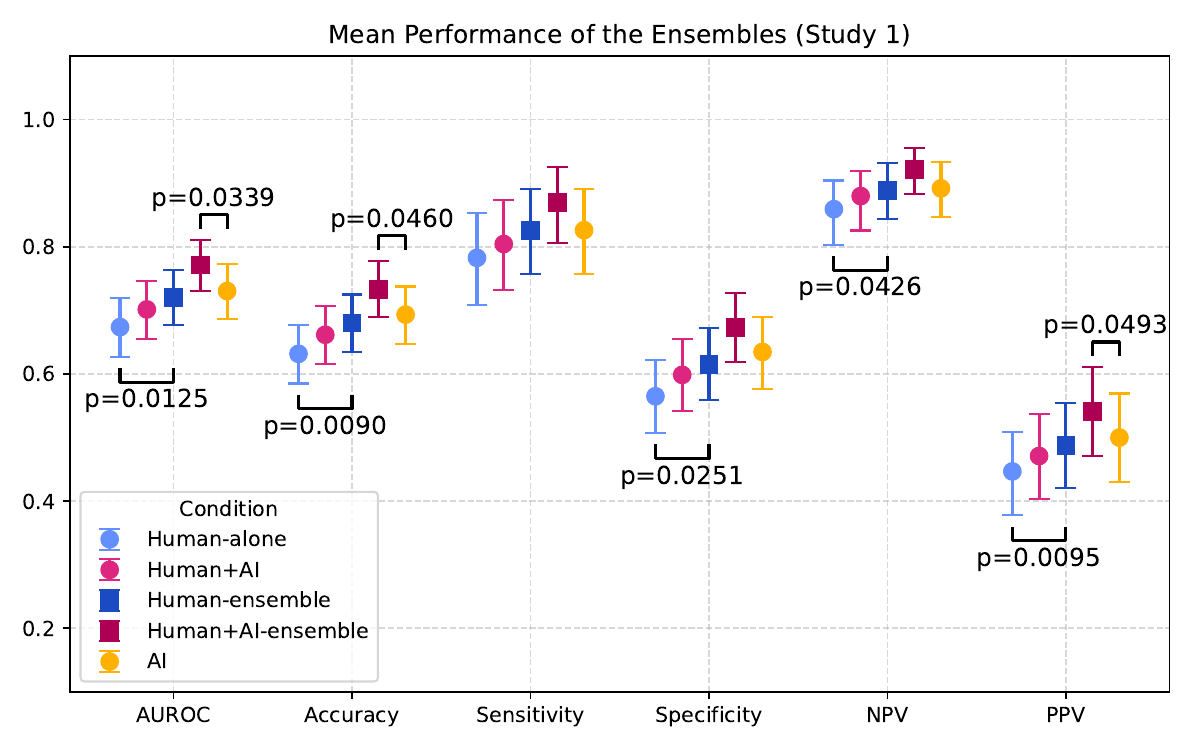}
    \caption{
        Mean performance of \halone, \hai, \hensem, \haiensem, and AI in Study 1.
        AUROC, accuracy, specificity, NPV, and PPV are significantly better in \hensem than in \halone.
        In \haiensem, AUROC, accuracy, and PPV are significantly better than that of AI.
    }
    \label{fig:ensemble}
\end{figure}

\para{Ensemble of human outperforms human but not AI, ensemble of Human+AI could outperform AI.} We compiled an ensemble of results from the human radiologists' predictions in \cref{tab:ensemble} and \cref{fig:ensemble}.
For each test case, we do a majority vote among the predictions from the eight radiologists.
If there is a tie among the radiologists, i.e. four cancer predictions versus four non-cancer predictions), we calculate the weighted prediction based on the radiologists' reported confidence.
Performance of \hensem is significantly improved over \halone, especially with precision/PPV increasing from $44.7\%$ to $48.7\%$ ($4\%$) and specificity rising from $56.5\%$ to $61.5\%$ ($5\%$).
This improvement closes the gap between humans and AI.
Moreover, \haiensem has the highest performance among all conditions, gaining significantly better AUROC ($0.771$), accuracy ($73.3\%$), and precision/PPV ($54.1\%$) than AI.
Sensitivity also reaches $87.0\%$, indicating a strong performance. 
This suggests that, with the help of AI, a group of experts can surpass either themselves or AI, achieving complementary performance.

\subsection{Behavioral Analysis on Human-AI collaboration (Q2)}

We now focus on the impact of different interventions, specifically the effect of performance feedback in Study 2 and the effect of providing AI assistance after humans have made their decisions.

\para{The different workflow does not significantly change human performance --- comparison of common-50 subset results of study 1 and 2.} In study 2, we share with each participant their own individual performance, the AI's performance, and their performance after reviewing AI predictions. 
A sample screenshot of the performance feedback provided to an individual radiologist is shown in \cref{fig:feedback}. 
To ensure radiologists understood their relative performance compared to the AI and whether AI assistance improved their results from Study 1, they were required to answer an attention check question before proceeding with the study.
We investigate how this performance feedback affects human decision making behavior, particularly whether they tended to incorporate AI advice more, less, or without significant change.
By learning about their past performance, the AI's performance, and the previous \hai team performance, radiologists were better informed before making new decisions in Study 2.

We hypothesized that radiologists would adjust their trust and reliance on AI if they realized that AI was more accurate overall. 
To test this, we analyze the performance of the 50 common test cases across study 1 and study 2. 
Despite the introduction of performance feedback, Human+AI team still does not surpass AI alone and achieves results that are relatively similar to or only slightly better than Human+AI in Study 1.
Moreover, there is no statistical significance in any of the metrics comparing \hai[2] with \hai[1]. 
As none of the metrics showed statistical significance, we defer the full details of the common-set results to Appendix \cref{tab:ensemble_common}. 
In conclusion, our findings suggest that performance feedback did not lead to significant improvements in the Human+AI accuracy.

\begin{figure}[t]
  \centering
  \begin{subfigure}[c]{0.6\textwidth}
    \centering
    \includegraphics[width=\textwidth]{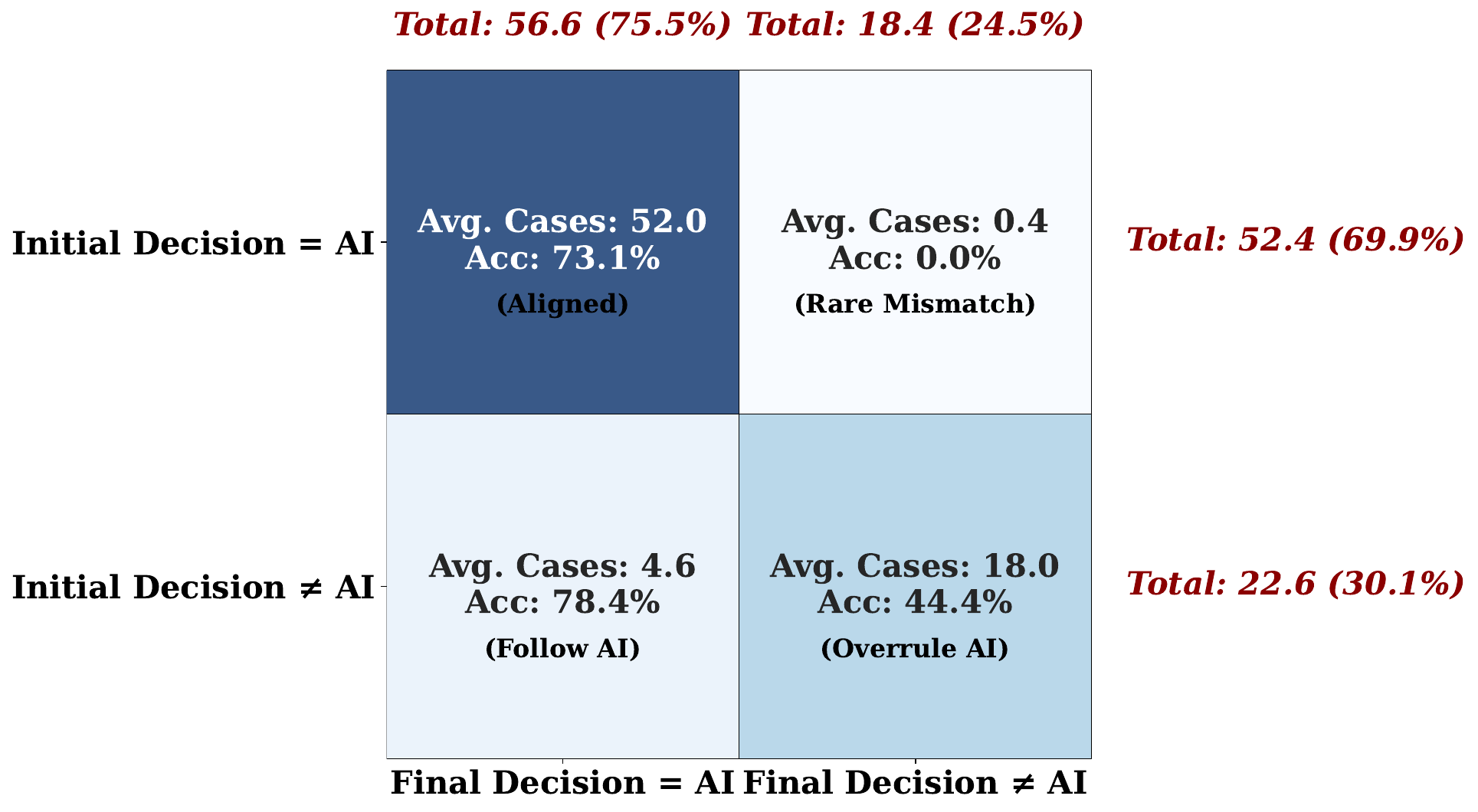}
    \caption{Study 1 (total cases n=75; radiologists N=8). Top-2 frequent groups are aligned and overrule AI. When there is a disagreement in the initial decision, radiologists are more likely to overrule AI predictions. However, Accuracy in the follow-AI group is higher than the `Aligned' and `Overrule AI' groups ($p=0.04^*$).
    }
    \label{fig:study1}
  \end{subfigure}
  \hspace{0.05\textwidth} %
  \begin{subfigure}[c]{0.3\textwidth}
    \centering
    \includegraphics[width=\textwidth]{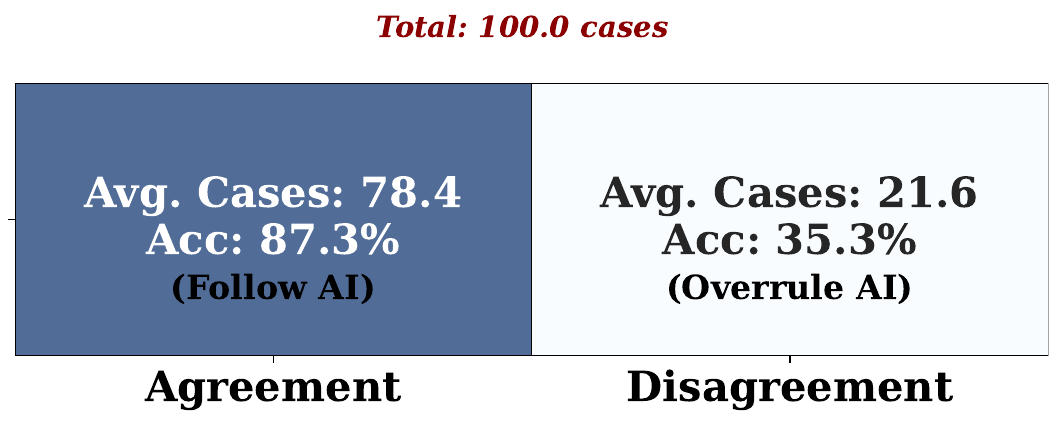}
    \caption{Study 2 (total cases n=100; radiologists n=8). Compared with study 1, radiologists are more likely to follow AI when the AI is shown directly without them making their own initial decision. Accuracy is also higher in the `follow AI' group compared with `overrule AI' ($p=0.00^*$). }
    \label{fig:study2}
  \end{subfigure}
  \caption{Comparison of Human-AI Decision Alignment and Accuracy. 
  Blue shading indicates frequency of cases for each scenarios; percentages showing diagnostic accuracy for scenario. 
  Accuracy is the highest in the follow-AI group for both studies. 
  }
  \label{fig:comparison}
\end{figure}

\para{Radiologists are reluctant in adopting AI assistance after they made their own independent diagnosis.}
In Study 1, radiologists first make diagnostic decisions before being shown the AI's predictions. This allows us to observe how likely they are to incorporate AI suggestions. 
The results indicate that radiologists tend to maintain their initial diagnostic decisions even when presented with contradicting AI predictions.
From \cref{fig:study1}, the initial agreement between human and AI is about 52.4 (69.9\%) vs. 22.6 (30.1\%). For 52.4 cases (initial agreement), human rarely changes their decision as their decision is confirmed by AI. 
When the AI disagrees with their initial assessment (22.6/75 average cases), radiologists change their diagnosis in only 4.6 (20.4\%) of cases.
This reluctance to revise initial decisions persists even in cases where their own accuracy is low (44.4\%), suggesting a significant barrier to incorporating AI assistance.

\begin{table}[t]
    \begin{center}
    \caption{Confidence score and time spent for the common 50-case subset. 
    }
    \label{tab:conf_time}
\resizebox{.9\textwidth}{!}{
    \begin{tabular}{lcccccc}
        \toprule
        & & \multicolumn{2}{c}{Study 1} & \multicolumn{2}{c}{Study 2} \\ \cmidrule(lr){3-4} \cmidrule(l){5-6}
        & Human & \hai[1] & P (\hai>Human)& \hai[2] & P (\hai>Human) \\ \midrule
        Confidence & \cella{3.34}{3.22}{3.47} & \cella{3.35}{3.23}{3.47} & $0.384$ & \cella{3.43}{3.31}{3.55} & $0.040^*$ \\ \midrule
        Time (s) & \cella{123.11}{110.47}{138.24} & \cella{144.65}{129.56}{161.96} & $0.000^*$ & \cella{115.89}{102.96}{130.05} & $0.225$ \\
        \bottomrule
    \end{tabular}
    }
    \end{center}
\end{table}

\begin{figure}[t]
  \centering
  \begin{subfigure}[b]{0.48\textwidth}
    \centering
    \includegraphics[width=\textwidth]{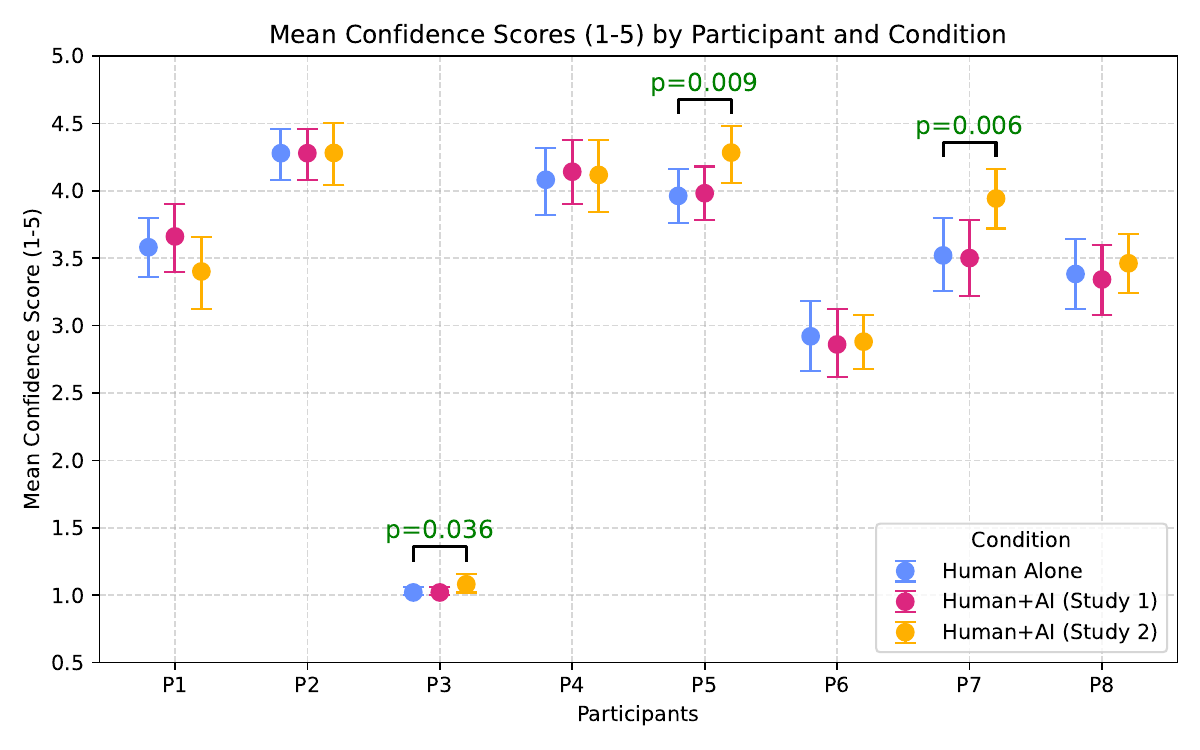}
    \caption{Mean confidence of each participant.}
  \end{subfigure}
  \begin{subfigure}[b]{0.48\textwidth}
    \centering
    \includegraphics[width=\textwidth]{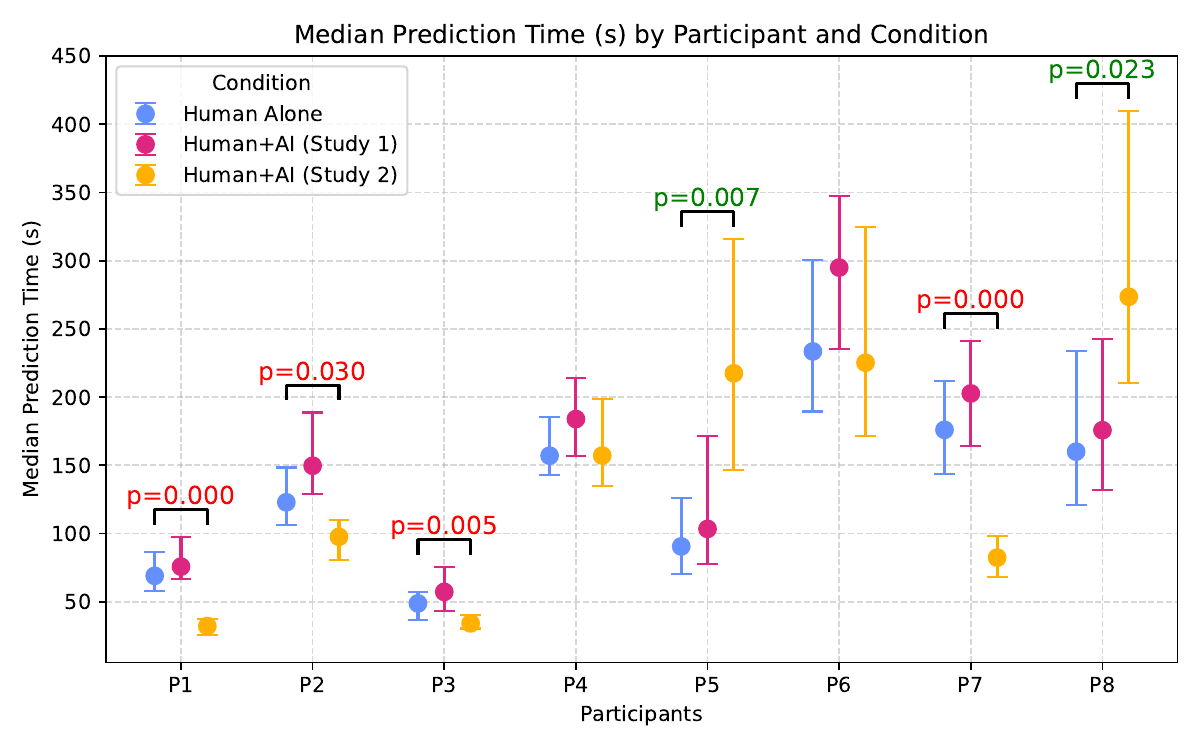}
    \caption{Median time spent of each participant.}
  \end{subfigure}
  \begin{subfigure}[t]{0.48\textwidth}
    \centering
    \includegraphics[width=\textwidth]{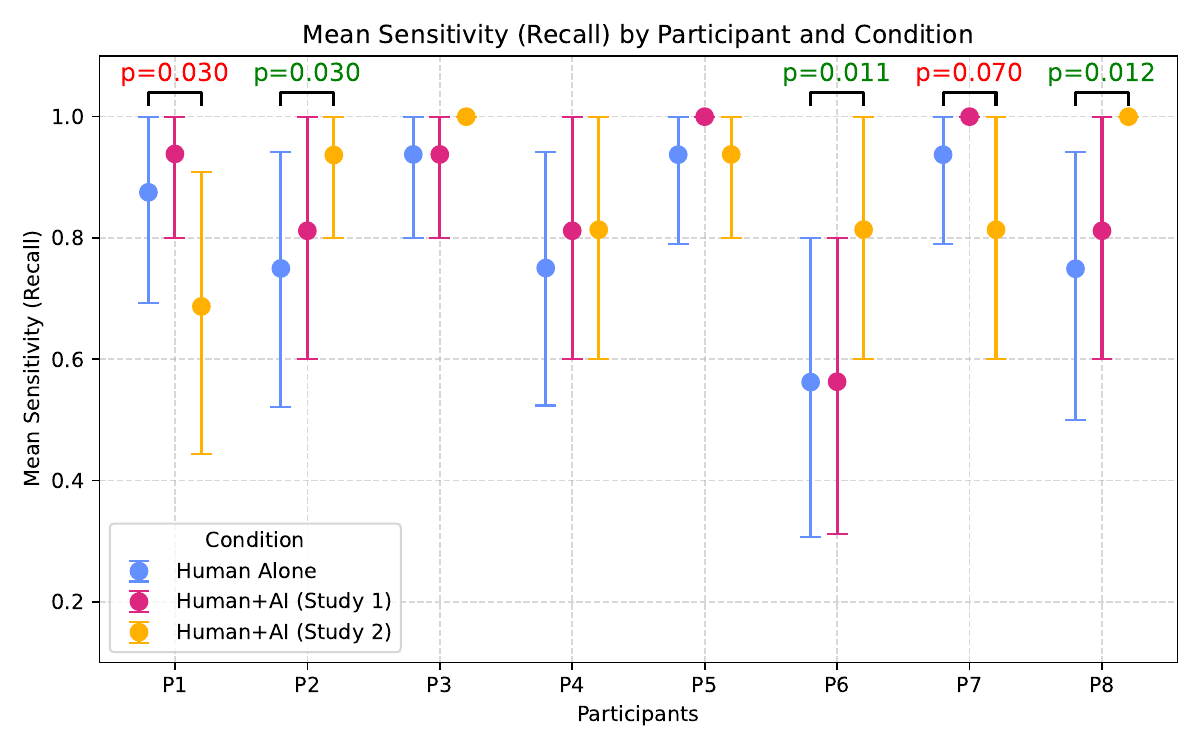}
    \caption{Mean sensitivity of each participant.}
  \end{subfigure}
  \begin{subfigure}[t]{0.48\textwidth}
    \centering
    \includegraphics[width=\textwidth]{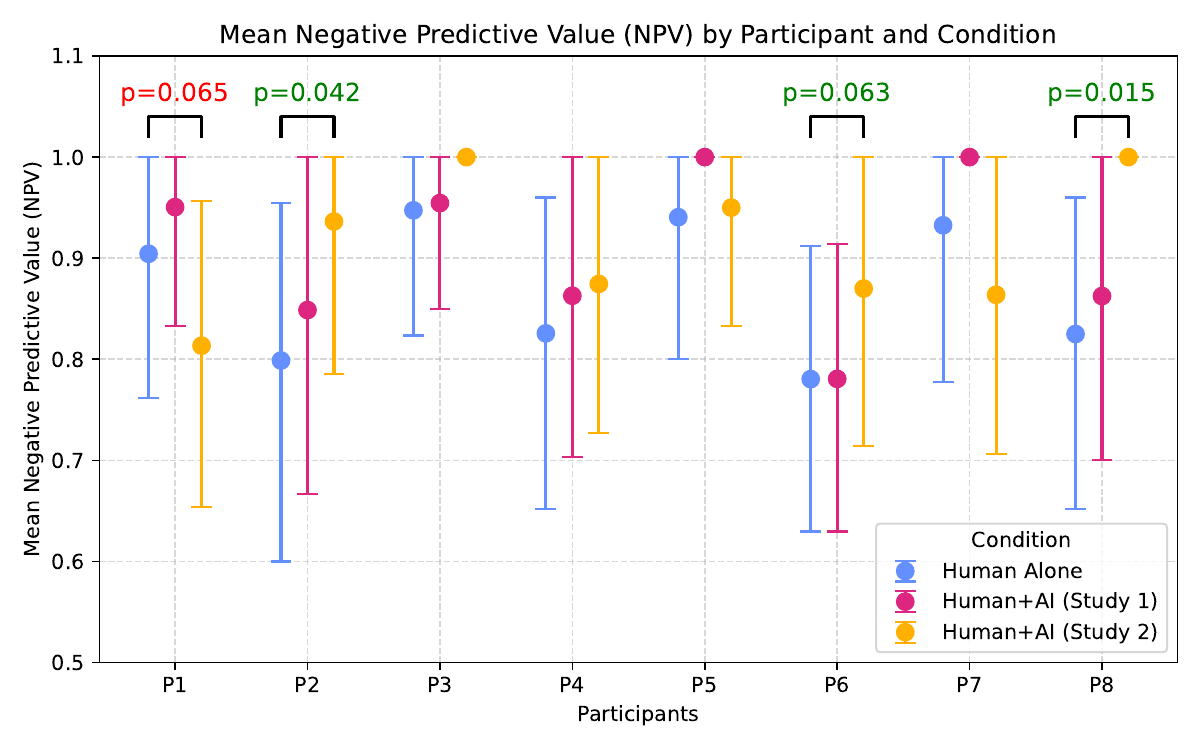}
    \caption{Mean negative predictive value of each participant.}
  \end{subfigure}
  \caption{
  The average confidence score, time spent, sensitivity, and NPV on the common 50-case subset for each participant.
  Statistics are calculated on data bootstrapped in the same way as the results in \cref{tab:common-cases}.
  Performance metrics other than sensitivity and NPV are excluded due to insignificance on the comparison between \halone and \hai[2].
  All significant comparisons are annotated with corresponding $p$-values with green indicating increasing and red indicating decreasing.
  Means are plotted with 95\% confidence Intervals.
  }
\end{figure}

\para{Upfront AI input and performance feedback increase AI adoption.}
In Study 2, performance feedback was shown to human radiologists at the very beginning of the study to help them gain a sense of their performance compared with AI from Study 1. When they diagnose each patient cases, AI predictions/assistance are shown upfront without them making their own initial diagnosis. 
As shown in \cref{fig:study2}, the results indicate that performance feedback and upfront AI assistance leads to higher rate of human-AI agreement (78.4\% ``follow AI'' vs. 75.5\% final human-AI agreement from study 1). 
In addition, ``follow AI'' group shows higher accuracy (87.3\%) compared with ``overrule AI'' group (35.3\%), as well as sensitivity (92.1\% vs. 36.6\%), and specificity (72.4\% vs. 34.8\%). For a complete results with more metrics, please refer to \cref{tab:study2_analysis} in the Appendix.
This slightly higher adoption rate, however, was insufficient to bridge the gap between \hai teams and AI significantly.

\para{Mixed Effects on Diagnostic Confidence and Time Efficiency.}
In addition to the diagnoses and annotations on the test cases, we also ask radiologists for a confidence score for each of their diagnoses on the case-level.
We design the confidence score to be on a scale of one to five (from ``Not certain at all'', ``Slightly certain'', ``Moderately certain'', ``Highly certain'', to ``Extremely certain'').
We observe no significant difference between the overall mean confidence scores of \halone and \hai[1].
However in \hai[2] radiologists report significantly higher confidence scores than in \halone ($p<0.05$), along with higher sensitivity and NPV as shown in \cref{tab:common-cases}.

We also tracked how long radiologists spent on each case in seconds. Because the \hai[1] diagnosis is an update of \halone, its recorded time includes the entire decision process from \halone. To mitigate outliers, we focus on median times: $123.11s$ for \halone, $144.65s$ for \hai[1], and $115.89s$ for \hai[2]. On average, radiologists used about 21 seconds more in \hai[1] (a statistically significant increase) and about 7 seconds less in \hai[2] (not statistically significant).
On the individual level, we did not observe a consistent ``time versus performance'' trade-off among all radiologists.  
Some spent less time and improved (P2) or maintained (P3, P5, P7) their diagnostic performance, while others lost sensitivity (P1). 
Others who took the same or more time either saw no change in performance (P5) or increased their sensitivity (P6, P8) or negative predictive values (P8).
These individual variations suggest that the relationship between processing time and diagnostic performance is complex and participant-specific.

\section{Conclusion}

While there is a growing interest in evaluating AI assistance with human decision makers, only a handful of previous works have attempted to evaluate AI systems directly with domain experts, and even fewer have achieved complementary performance or investigated human behavior. 
We contribute a comprehensive study with domain experts about how a clinical AI tools might be integrated in practice with two realistic design of workflows. Our findings suggest that while human-AI teams consistently outperform humans alone, they still underperform compared to AI due to under-reliance.
More importantly, we look beyond accuracy and investigate human behavioral patterns in human-AI interaction. 
Even when domain experts are informed about their performance, the gap to AI performance, and their previous AI-assisted performance, it remains challenging for them to effectively calibrate their reliance and trust in AI tools. 
While complementary performance falls short in our work---as in previous works---our results on the ensemble performance of human-AI teams are promising.
This highlights exciting opportunities to improve human-AI decision-making.

\para{Limitations.} 
Several issues remain unresolved and present opportunities for future research. 
While our study show that upfront AI assistance can encourage greater adoption among radiologists, it remains unclear what factors positively contribute to complementary performance.
Additionally, our research is limited to particular clinical setting and disease, which may not be generalizable to other domains or environments. 
Despite these limitations, we hope that our study will inspire and support the broader research community to further investigate the complexities of human-AI decision-making in relevant real-world tasks.

% \newpage
\bibliographystyle{ACM-Reference-Format}
\bibliography{reference}

% \newpage
\appendix

\section{Model Impementation Details}
\label{appen:model}

\para{Training configurations}
We use the established nnU-Net implementation\footnote{\url{https://github.com/DIAGNijmegen/picai_baseline}} for image segmentation.
The framework was configured to handle dataset preprocessing, augmentation, and training pipeline generation automatically. The training process utilized a batch size of 8 and a learning rate of $0.001$, optimized using the AdamW optimizer. Training was performed over 1000 epochs on one NVIDIA A40 GPU. nnU-Net's default data augmentation techniques, such as random cropping, flipping, and intensity scaling, were employed to improve generalization. For lesion-level prediction, we set the threshold to 0.5. The framework's automatic hyperparameter tuning ensured optimal performance, and we monitored model training using AUROC and average precision on the validation set.
A detailed performance is shown in table~\cref{tab:AI-results}.

\begin{table}[ht]
\begin{tabular}{@{}lcccccccc@{}}
\toprule
                   & \multicolumn{4}{c}{Training (n=1211)}        & \multicolumn{4}{c}{Testing (n=200)}         \\         \cmidrule(lr){2-5} \cmidrule(l){6-9}
                   & AUROC  & AP     & Accuracy & F1     & AUROC  & AP     & Accuracy & F1     \\
                   \midrule
Per-patient & 0.910 & 0.737 & 0.847   & 0.725 & 0.799 & 0.624 & 0.735    & 0.644 \\
Per-lesion       & 0.940 & 0.682 & 0.948   & 0.664 & 0.824 & 0.484 & 0.911   & 0.531 \\ \bottomrule
\end{tabular}
\label{tab:AI-results}
\caption{AI model performance.}
\end{table}

\section{Demographics}
\label{app:human_demographics}
We recruit 8 practicing radiologists, aged 29 to 52 years (mean: 38.4 years). Respondents were primarily from the United States (n=4), Turkey (n=3), and Italy (n=1). Most participants (n=5) reported advanced or expert-level experience with prostate MRI, whilte the others reported intermediate (n=2). One participant did not answer this question.  

\section{Exit Survey Results}

\subsection*{Study 1 Results}
In Study 1, participants were highly familiar with the AI tool (mean familiarity: 5/5), though its accuracy received a lower mean rating of 2.4/5. Usefulness and trust in the system were rated moderately, both averaging 3/5. 
In open-ended feedback, practitioners reported that the AI tool was most helpful in ambiguous cases and increased confidence in detecting lesions in challenging locations such as the anterior, apical, and transition zones. Concerns included oversensitivity in non-cancerous areas and missed lesions, with suggestions for improvement focusing on providing malignancy probability scores, separate reporting of T2 and DWI/ADC scores, and better performance in transitional zone lesions.

\subsection*{Study 2 Results}
In Study 2, the AI tool's helpfulness was rated moderately (mean: 2.9/5), with accuracy ratings remaining low to moderate (mean: 2.1/5). Trust in the AI also averaged 2.5/5. Despite moderate satisfaction, respondents expressed a high likelihood of future AI use (mean: 3.75/5). In open-ended feedback, the AI was perceived as useful in ambiguous cases, with one practitioner noting it reinforced decisions to call studies negative. 
They also pointed out key challenges such as poor performance in transitional zone lesions, overreliance on diffusion restriction, and limitations in segmenting prostate versus non-prostate tissue. Participants' recommendations for improvement included adopting the PI-RADS classification system, enhancing segmentation capabilities, and improving detection of small lesions. Image quality issues were a significant limitation, with practitioners noting that humans outperform AI in evaluating non-diagnostic images, particularly for diffusion-weighted imaging.

\section{Fine-grained analysis}
\label{app:fine_grained}

\cref{tab:study1_analysis} and \cref{tab:study2_analysis} provide an overview of the subgroup analysis of human-AI agreement and disagreement in Studies 1 and 2, respectively. The results indicate that performance metrics are significantly better in subgroups where human and AI decisions align compared to those with disagreement.

For a detailed breakdown, individual-level performance for the different agreement and disagreement subgroups is presented. In Study 1, the results are available in \cref{tab:study1-fine-1}, \cref{tab:study1-fine-2}, \cref{tab:study1-fine-3}, and \cref{tab:study1-fine-4}, each focusing on specific subcategories of agreement or disagreement. Similarly, Study 2 individual-level results are provided in \cref{tab:study2-fine}, offering finer granularity of the analysis.

\begin{table}[t]
\centering
\caption{Study 1 fine-grained subgroup performance.}
\resizebox{\textwidth}{!}{
\begin{tabular}{@{}lcccccccccc@{}}
\toprule
\textbf{Condition}         & \textbf{Avg (\#)} & \textbf{Total} & \textbf{Correct} & \textbf{TP} & \textbf{FP} & \textbf{TN} & \textbf{FN} & \textbf{Acc (\%)} & \textbf{Sen (\%)} & \textbf{Spc (\%)} \\ \midrule
Initial=AI, final=AI         & 52.0         & 416            & 304              & 122         & 99          & 182         & 13          & 73.1              & 90.4              & 64.8              \\
Initial=AI, final$\neq$AI    & 0.4          & 3              & 0                & 0           & 0           & 0           & 3           & 0.0               & 0.0               & N/A               \\
Initial$\neq$AI, final=AI    & 4.6          & 37             & 29               & 10          & 5           & 19          & 3           & 78.4              & 76.9              & 79.2              \\
Initial$\neq$AI, final$\neq$AI & 18.0         & 144            & 64               & 16          & 63          & 48          & 17          & 44.4              & 48.5              & 43.2              \\ \bottomrule
\end{tabular}
}
\label{tab:study1_analysis}
\end{table}

\begin{table}[t]
\centering
\caption{Study 2 fine-grained subgroup performance.}
\resizebox{\textwidth}{!}{
\begin{tabular}{@{}lcccccccccc@{}}
\toprule
\textbf{Condition}                & \textbf{Avg (\#)} & \textbf{Total} & \textbf{Correct} & \textbf{TP} & \textbf{FP} & \textbf{TN} & \textbf{FN} & \textbf{Acc (\%)} & \textbf{Sen (\%)} & \textbf{Spc (\%)} \\ \midrule
Human $\neq$ AI prediction        & 21.6             & 173            & 61               & 15          & 86          & 46          & 26          & 35.3              & 36.6              & 34.8              \\
Human $=$ AI prediction           & 78.4             & 627            & 496              & 198         & 114         & 298         & 17          & 79.1              & 92.1              & 72.4              \\ \bottomrule
\end{tabular}}
\label{tab:study2_analysis}
\end{table}

\begin{table}[]
\caption{
Study 1: Cases where human agreed with AI and decision was kept.
}
\begin{tabular}{@{}cccccccc@{}}
\toprule
\textbf{Username} & \textbf{Total Cases} &  \textbf{Correct} & \textbf{TP} & \textbf{FP} & \textbf{TN} & \textbf{FN} & \textbf{Accuracy} \\
\midrule
P1   & 53             & 40               & 17          & 12          & 23          & 1           & 75.5\%     \\
P2  & 46             & 33               & 15          & 13          & 18          & 0           & 71.7\%     \\
P3         & 67             & 47               & 19          & 17          & 28          & 3           & 70.1\%     \\
P4   & 51             & 37               & 14          & 12          & 23          & 2           & 72.5\%     \\
P5    & 51             & 37               & 18          & 12          & 19          & 2           & 72.5\%     \\
P6    & 46             & 36               & 9           & 8           & 27          & 2           & 78.3\%     \\
P7      & 50             & 35               & 16          & 14          & 19          & 1           & 70.0\%    \\
P8 & 52             & 39               & 14          & 11          & 25          & 2           & 75.0\%     \\ \bottomrule
\end{tabular}
\label{tab:study1-fine-1}
\end{table}

\begin{table}[]
\caption{Study 1: Cases where human agreed but AI initially but still changed decision against AI.}
\begin{tabular}{@{}cccccccc@{}}
\toprule
\textbf{Username} & \textbf{Total Cases} &  \textbf{Correct} & \textbf{TP} & \textbf{FP} & \textbf{TN} & \textbf{FN} & \textbf{Accuracy} \\
\midrule
 P6       & 3                    & 0                    & 0                    & 0                    & 0                    & 3                    & 0.00\%                  \\
                \bottomrule
\end{tabular}
\label{tab:study1-fine-2}
\end{table}

\begin{table}[]
\caption{
Study 1: cases where human disagreed with AI but kept original decision.}
\begin{tabular}{@{}cccccccc@{}}
\toprule
\textbf{Username} & \textbf{Total Cases} &  \textbf{Correct} & \textbf{TP} & \textbf{FP} & \textbf{TN} & \textbf{FN} & \textbf{Accuracy} \\
\midrule
P1   & 20             & 9                & 2           & 10          & 7           & 1           & 45.0\%         \\
P2  & 23             & 10               & 4           & 10          & 6           & 3           & 43.5\%     \\
P3         & 2              & 1                & 0           & 1           & 1           & 0           & 50.0\%          \\
P4   & 18             & 8                & 2           & 7           & 6           & 3           & 44.4\%     \\
P5    & 20             & 9                & 2           & 11          & 7           & 0           & 45.0\%         \\
P6    & 22             & 12               & 2           & 5           & 10          & 5           & 54.5\%     \\
P7      & 18             & 6                & 2           & 11          & 4           & 1           & 33.3\%     \\
P8 & 21             & 9                & 2           & 8           & 7           & 4           & 42.9\%     \\ \bottomrule
\end{tabular}
\label{tab:study1-fine-3}
\end{table}

\begin{table}[]
\caption{
Study 1: cases where human disagreed with AI but followed AI advice.}
\begin{tabular}{@{}cccccccc@{}}
\toprule
\textbf{Username} & \textbf{Total Cases} &  \textbf{Correct} & \textbf{TP} & \textbf{FP} & \textbf{TN} & \textbf{FN} & \textbf{Accuracy} \\
\midrule
P1   & 2              & 1                & 1           & 0           & 0           & 1           & 50.0\%          \\
P2  & 6              & 6                & 1           & 0           & 5           & 0           & 100.0\%            \\
P3         & 6              & 4                & 0           & 1           & 4           & 1           & 66.7\%     \\
P4   & 6              & 5                & 2           & 1           & 3           & 0           & 83.3\%     \\
P5    & 4              & 4                & 1           & 0           & 3           & 0           & 100.0\%            \\
P6    & 4              & 3                & 2           & 1           & 1           & 0           & 75.0\%         \\
P7      & 7              & 5                & 2           & 1           & 3           & 1           & 71.4\%     \\
P8 & 2              & 1                & 1           & 1           & 0           & 0           & 50.0\%          \\ \bottomrule
\end{tabular}
\label{tab:study1-fine-4}
\end{table}

\begin{table}[]
\caption{Finegrained analysis for Study 2: (1) When Human disagrees with AI, human are prone to errors (accuracy is lower than 50\%); (2) Human is better at identifying AI false positives than identifying false negatives, i.e., humans are better at catching AI’s false alarms than its missed cases.}
\begin{tabular}{@{}cccccccc@{}}
\toprule
\textbf{Username} & \textbf{\#Disagreements} &  \textbf{Correct} & \textbf{TP} & \textbf{FP} & \textbf{TN} & \textbf{FN} & \textbf{Accuracy} \\
\midrule
P1   & 28                    & 11      & 1  & 10 & 10 & 7  & 39.3\% \\
P2  & 27                    & 6       & 2  & 19 & 4  & 2  & 22.2\% \\
P3         & 11                    & 3       & 3  & 8  & 0  & 0  & 27.3\% \\
P4   & 26                    & 11      & 1  & 11 & 10 & 4  & 42.3\% \\
P5    & 18                    & 7       & 1  & 9  & 6  & 2  & 38.9\% \\
P6    & 20                    & 8       & 2  & 6  & 6  & 6  & 40.0\%      \\
P7      & 20                    & 6       & 2  & 11 & 4  & 3  & 30.0\%      \\
P8 & 23                    & 9       & 3  & 12 & 6  & 2  & 39.1\% \\ \bottomrule
\end{tabular}
\label{tab:study2-fine}
\end{table}

\section{Ensemble on Common-50 Cases}

\cref{tab:ensemble-common-cases} presents a detailed performance comparison among AI, Human, Human-ensemble, Human+AI, and Human+AI ensemble (Study 1 and Study 2) for the common 50-case subset. While the results highlight that the Human-ensemble consistently outperforms individual human performance, the advantage of any ensemble method over AI alone is less significant.

\begin{table}[t]
    \begin{center}
    \caption{Performance comparison between AI, Human, Human-ensemble, Human+AI, and human+AI ensemble (study 1 and 2) for the common 50-case subset. 
    }
    \label{tab:ensemble_common}
\resizebox{\textwidth}{!}{
    \begin{tabular}{lccccccccc}
        \toprule
        && \multicolumn{5}{c}{Study 1} & \multicolumn{3}{c}{Study 2}\\
        \cmidrule(lr){3-7} \cmidrule(l){8-10}
        & AI & Human & Human-ensemble & Human+AI & H+AI ensemble & \tabcell{P (\hensem>Human)\\P (H+AI ensemble>AI)} & Human+AI & H+AI ensemble & P (H+AI ensemble>AI)  \\ \midrule
        AUROC & \cella{0.763}{0.727}{0.797} & \cella{0.675}{0.630}{0.719} & \cella{0.732}{0.690}{0.771} & \cella{0.711}{0.668}{0.752} & \cella{0.778}{0.741}{0.812} & $0.004^*$/$0.265$ & \cella{0.708}{0.666}{0.748} & \cella{0.763}{0.726}{0.798} & $0.112$ \\ \midrule
        Accuracy & \cellb{70.0\%}{0.657}{0.745}{35}{50} & \cellb{62.5\%}{0.578}{0.672}{31}{50} & \cellb{68.0\%}{0.635}{0.725}{34}{50} & \cellb{65.7\%}{0.610}{0.703}{33}{50} & \cellb{72.0\%}{0.675}{0.762}{36}{50} & $0.004^*$/$0.216$ & \cellb{64.7\%}{0.600}{0.693}{32}{50} & \cellb{70.0\%}{0.655}{0.745}{35}{50} & $0.229$ \\ \midrule
        Sensitivity (Recall) & \cellb{93.8\%}{0.892}{0.976}{15}{16} & \cellb{81.2\%}{0.741}{0.878}{13}{16} & \cellb{87.5\%}{0.814}{0.929}{14}{16} & \cellb{85.9\%}{0.797}{0.917}{14}{16} & \cellb{93.8\%}{0.892}{0.976}{15}{16} & $0.028^*$/$0.495$ & \cellb{87.5\%}{0.815}{0.929}{14}{16} & \cellb{93.8\%}{0.892}{0.976}{15}{16} & $0.050$ \\ \midrule
        Specificity & \cellb{58.8\%}{0.530}{0.646}{20}{34} & \cellb{53.7\%}{0.477}{0.595}{18}{34} & \cellb{58.8\%}{0.529}{0.646}{20}{34} & \cellb{56.2\%}{0.504}{0.620}{19}{34} & \cellb{61.8\%}{0.559}{0.675}{21}{34} & $0.027^*$/$0.197$ & \cellb{54.0\%}{0.482}{0.599}{18}{34} & \cellb{58.8\%}{0.528}{0.647}{20}{34} & $0.498$ \\ \midrule
        NPV & \cellb{95.2\%}{0.918}{0.982}{20}{21} & \cellb{87.0\%}{0.804}{0.909}{18}{21} & \cellb{90.9\%}{0.864}{0.949}{20}{22} & \cellb{90.8\%}{0.846}{0.938}{19}{21} & \cellb{95.5\%}{0.921}{0.983}{21}{22} & $0.012^*$/$0.467$ & \cellb{91.4\%}{0.854}{0.945}{18}{20} & \cellb{95.2\%}{0.919}{0.982}{20}{21} & $0.051$ \\ \midrule
        PPV (Precision) & \cellb{51.7\%}{0.453}{0.581}{15}{29} & \cellb{45.5\%}{0.389}{0.517}{13}{29} & \cellb{50.0\%}{0.435}{0.566}{14}{28} & \cellb{48.2\%}{0.416}{0.545}{14}{29} & \cellb{53.6\%}{0.470}{0.602}{15}{28} & $0.005^*$/$0.214$ & \cellb{47.4\%}{0.410}{0.537}{14}{30} & \cellb{51.7\%}{0.452}{0.582}{15}{29} & $0.236$ \\
        \bottomrule
    \end{tabular}
    }
    \label{tab:ensemble-common-cases}
    \end{center}
\end{table}

\section{More Screenshots on User Interface Design}
\label{sec:ui-design}

\begin{figure}[ht]
    \centering
    \includegraphics[width=.9\linewidth]{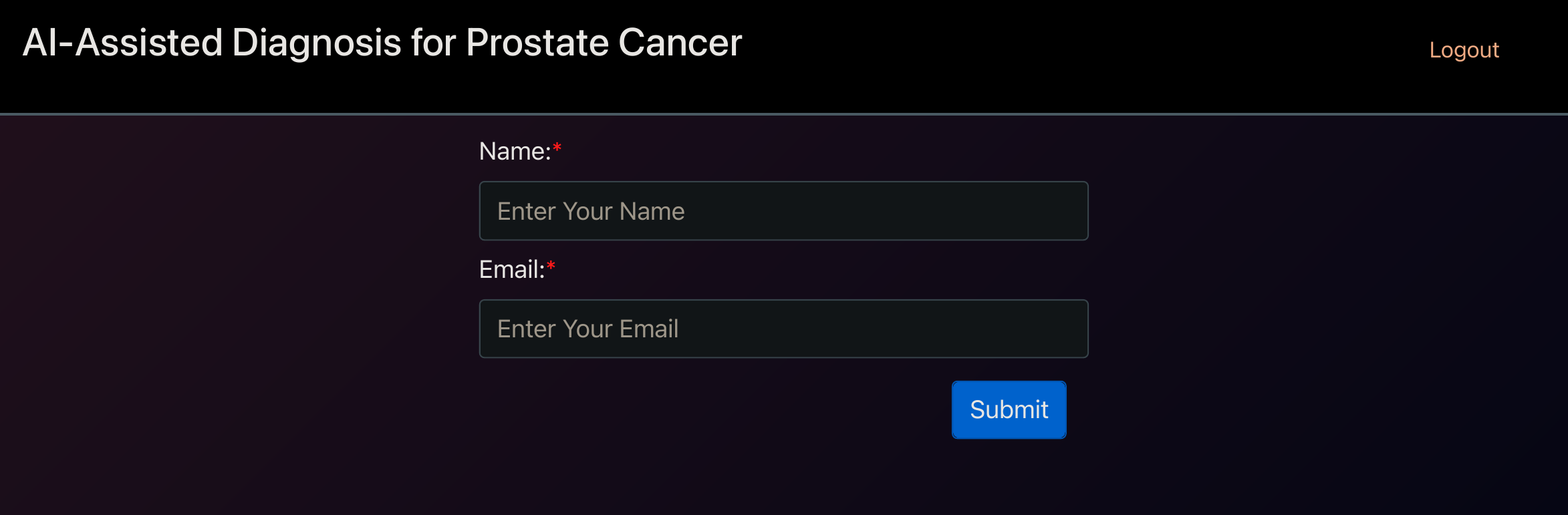}
    \caption{Login page.}
    \label{fig:login}
\end{figure}

\begin{figure}[ht]
    \centering
    \includegraphics[width=\linewidth]{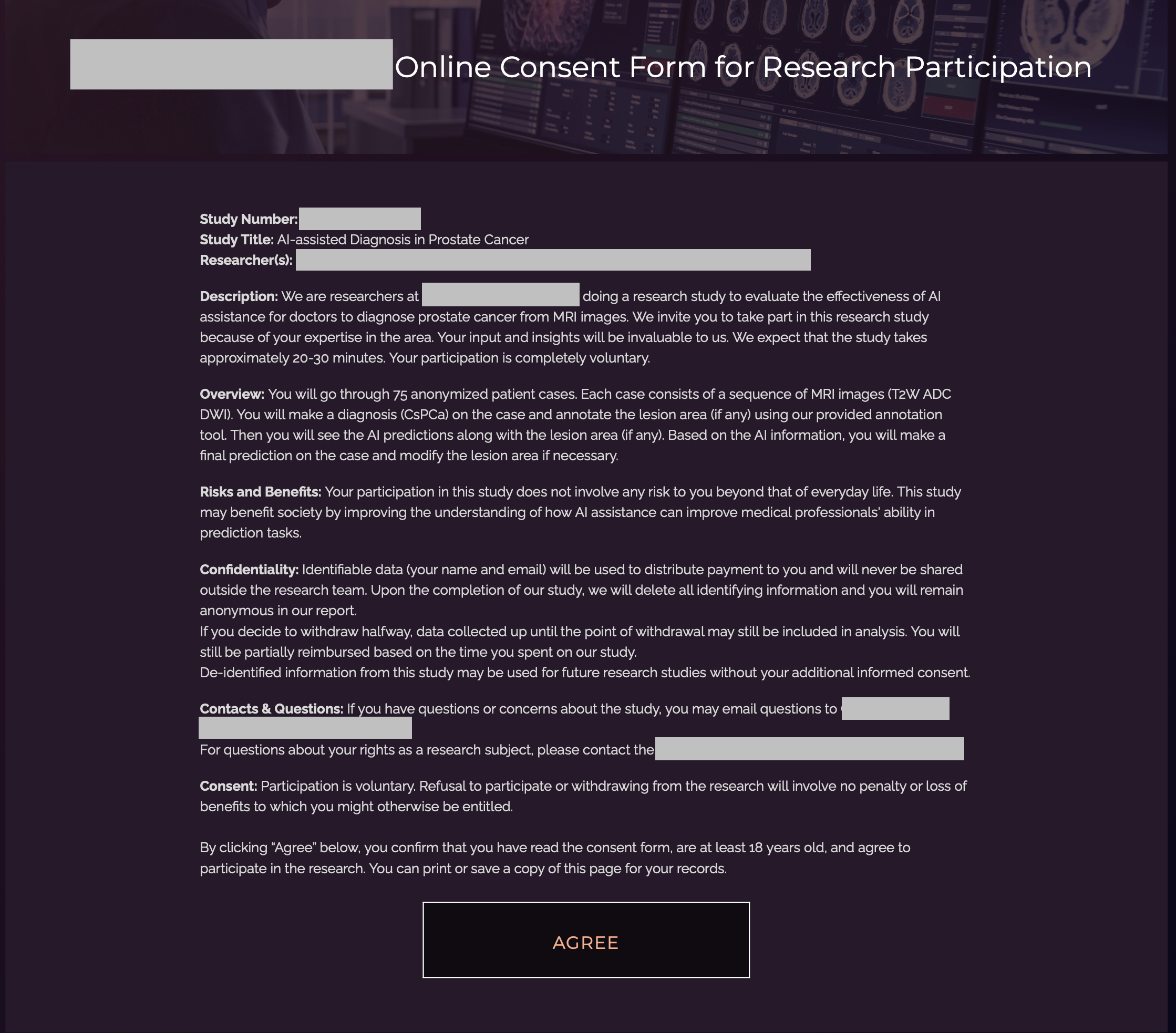}
    \caption{Consent page.}
    \label{fig:consent-form}
\end{figure}

\begin{figure}[ht]
    \centering
    \includegraphics[width=\linewidth]{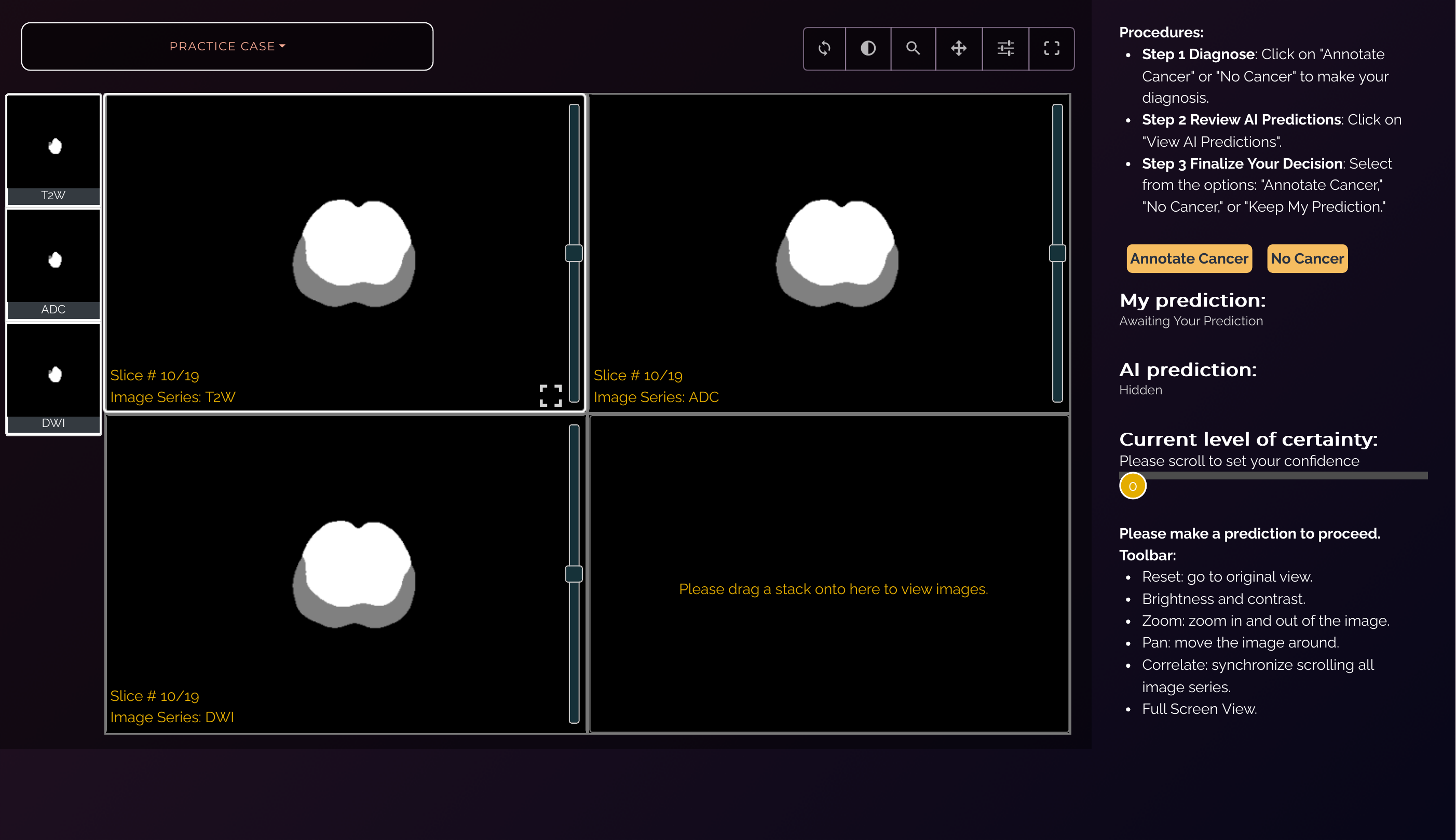}
    \caption{Toy demonstration example page.}
    \label{fig:demo}
\end{figure}

\begin{figure}
    \centering
    \includegraphics[width=.55\linewidth]{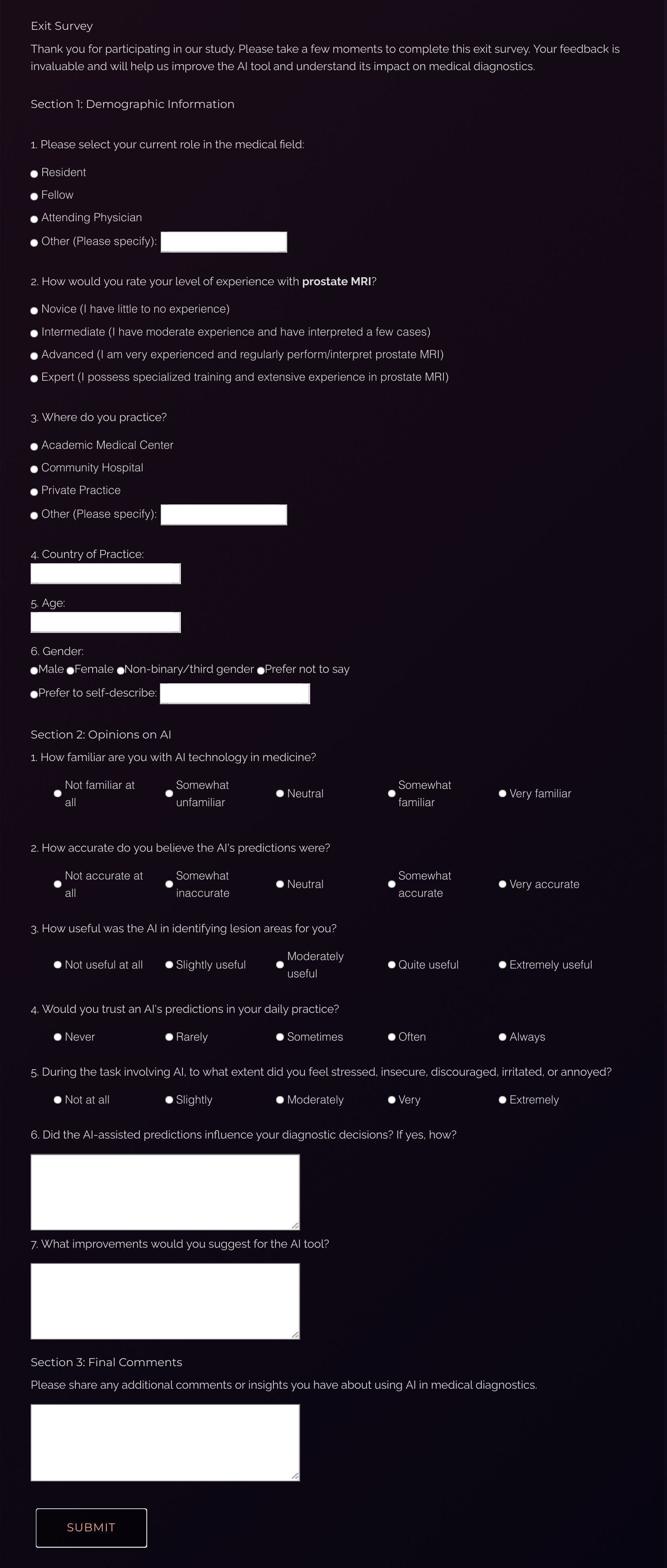}
    \caption{Exit survey for study 1.}
    \label{fig:exit-survey-1}
\end{figure}

\begin{figure}
    \centering
    \includegraphics[width=.55\linewidth]{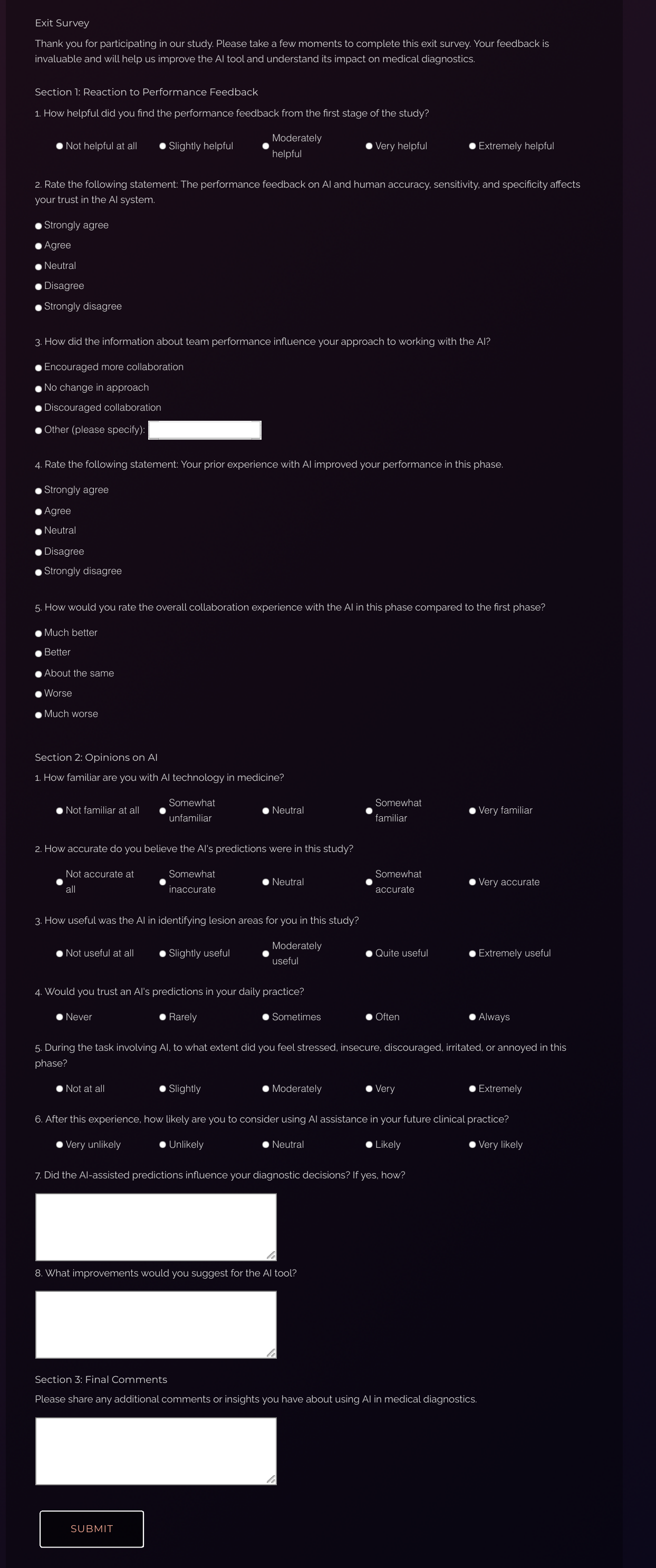}
    \caption{Exit survey for study 2.}
    \label{fig:exit-survey-2}
\end{figure}

We show screenshots of a login page (\cref{fig:login}), a consent form (\cref{fig:consent-form}), a toy demonstration example page (\cref{fig:demo}), and two exit surveys (\cref{fig:exit-survey-1}, \cref{fig:exit-survey-2}) for study 1 and study 2 respectively.

\end{document}